\documentclass[12pt,preprint]{aastex}

\begin{document}

\title{A new formation model
for $\omega$ Centauri: the crossroad of astrophysical processes}

\author{Kenji Bekki} 
\affil{
ICRAR,
M468,
The University of Western Australia
35 Stirling Highway, Crawley
Western Australia, 6009, Australia
}

\and

\author{Takuji Tsujimoto}
\affil{
National Astronomical Observatory of Japan, Mitaka, Tokyo 181-8588, Japan}

\begin{abstract}
We investigate the formation processes of the Galactic globular cluster (GC)
$\omega$ Cen
with multiple stellar populations based on our original hydrodynamical
simulations with chemical enrichment by Type II supernovae (SNe II), 
asymptotic giant
branch (AGB) stars, and neutron star mergers  (NSMs).
Multiple stellar populations with a wide range of [Fe/H]
can be formed from rather massive and compact molecular cloud
with a mass of $\approx 2 \times 10^7 M_{\odot}$ 
in the central region of its dwarf galaxy within less than a few hundred Myr.
Gas ejected from  SNe II and AGB stars can mix well to form 
new stars with higher He abundances ($Y$) and higher [Fe/H].
The He-rich stars are strongly concentrated in the GC's central region
so that the GC can show a steep negative gradient of $Y$. 
Relative ratios of light elements
to Fe show bimodal distributions for a given [Fe/H]
owing to star formation from original gas and AGB ejecta.
[La/Fe] and [Ba/Fe] can rapidly
increase until [Fe/H]$\sim -1.5$  and then decrease 
owing to Fe ejection from SNe II. 
Although AGB ejecta can be almost fully retained 
in intra-cluster medium,
NSM ejecta can be retained only partially.
This difference in the retention capability is responsible for
the observed unique [Eu/Fe]$-$[Fe/H] and [La/Eu]$-$[Fe/H] 
relations in $\omega$ Cen.
The observed [O/Na]$-$[Fe/H]
relation and radial [Fe/H] gradient  are yet to be well reproduced
in the present model.
We briefly discuss how the results change for different yields of AGB stars
and SNe II.
\end{abstract}


\section{Introduction}

The formation of globular clusters (GC) with multiple stellar
populations
is the crossroad of various astrophysical processes. It involves,
for example,  
long-term dynamical evolution driven by two-body
relaxation (e.g., Vesperini et al. 2010),
dynamical influences of the Galactic disk 
(e.g., Gnedin et al. 1999), 
mixing of intra-cluster medium
(ICM) with gas ejected from
fast-rotating massive stars (e.g., Prantzos \& Charbonel 2006)
and from asymptotic giant branch (AGB) stars (e.g., D'Antona \& Ventura 2002),
secondary star formation within existing dense  stellar systems 
(e.g., D'Ercole et al. 2008),
enrichment of $r$-process elements due to
efficient retention of gaseous  ejecta from neutron star mergers (NSMs)
in high-density ICM (Bekki \& Tsujimoto 2017;
BT17),
and stellar nucleosynthesis  (e.g., Renzini et al. 2015).
It is crucial for any theoretical study of GC formation
to understand the relative importance of each of these processes
in the formation of their multiple stellar populations.
However,  no theory has so far explained the observed
various  properties of 
their multiple stellar populations
 in a self-consistent manner (e.g., see Bastian \& Lardo 2018 for  a recent
review).

The Galactic GC $\omega$ Cen is a unique laboratory to study
physical processes related to the origin 
of multiple stellar populations of GCs, 
because there are so many previous observational studies on 
chemical and dynamical  properties
of $\omega$ Cen.
It is observed to have
characteristic features such as  its large mass
and flattened shape (e.g., Meylan 1987; Meylan et al. 1995),
possibly different kinematics in different stellar populations
(e.g., Norris et al. 1997; Bellini et al. 2018),
large metallicity dispersion (e.g., Freeman \& Rodgers 1975;
Norris et al. 1996),
different spatial distributions among multiple stellar
populations (e.g., Pancino et al. 2000;
Ferraro et al. 2002),
possible age spreads among the populations (e.g.,
Lee et al. 1999; Smith et al. 2000, S00; Marino et al. 2011);
retrograde orbit with respect to the Galactic rotation (e.g., Dinescu et al. 1999),
double main sequence in the color magnitude diagram (e.g.,
Anderson 1997; Bedin et al. 2004),
large spread in helium abundance $Y$ (e.g., Piotto et al. 2005),
and radial gradient of $Y$ (e.g., Sollima et al. 2007).
Recent observations of color magnitude relations of $\omega$ Cen
have identified 15 subpopulations with possibly different chemical
abundances in $\omega$ Cen (e.g., Bellini et al. 2017). Different  theoretical
models for $\omega$ Cen  have tried to provide physical
explanations for these unique characteristics of $\omega$ Cen
(e.g., Carraro \& Lia 2000;  Bekki \& Freeman  2003, BF03;  Tsujimoto et al. 2003).

One of promising scenarios for the formation of $\omega$ Cen is that 
it had been the nucleus of a dwarf galaxy that was completely
destroyed by the early formation phase of the Galaxy (e.g., Freeman 1993). 
Dynamical evolution of a nucleated dwarf galaxy into a naked
nucleus (i.e., $\omega$ Cen) has been investigated
in details by several authors (e.g., BF03;
Mizutani \& Chiba 2003;
Ideta et al. 2005).
For example, BF03 demonstrated that (i) a large amount of more metal-rich gas
can be transferred into the nucleus of a gas-rich nucleated dwarf
and (ii) the nucleus can therefore  have multiple generation of stars with
different metallicities.
Previous observations discovered possible tidal debris from $\omega$ Cen
(e.g.,  Wylie de-Boer et al. 2010) and
recent observations of the Galactic halo stars based on Gaia  data
have revealed the evidence of stellar streams kinematically
associated with
$\omega$ Cen (e.g., Myeong et al. 2018; Ibata et al. 2019). 
These observations
imply that  the above formation
scenario of $\omega$ Cen is promising.

Although dynamical origins of $\omega$ Cen have been extensively
discussed so far,  its chemical properties have not been modeled
in detail.
Ikuta \& Arimoto (2000) adopted the self-enrichment 
scenario of $\omega$ Cen and thereby investigated how the initial
mass function of stars (IMF) and gaseous outflow and inflow
during the early evolution of $\omega$ Cen
can control the distribution of [Ca/H] in $\omega$ Cen.
Using one-zone chemical evolution models of $\omega$ Cen,
Romano et al. (2007) tried to reproduce the observed abundance
distribution function (ADF), age-metallicity relation (AMR),
and [Fe/H]$-{\rm [\alpha/Fe]}$ relation. They modeled a long-term
($\sim 3$ Gyr) star formation history in order to reproduce the
possibly decreasing ${\rm [\alpha/Fe]}$ and increasing [Cu/Fe] with increasing
[Fe/H] (for $-2 \le {\rm [Fe/H]} \le -0.3$) in $\omega$ Cen.
Marcolini et al. (2007) investigated the long-term ($\sim 1.5$ Gyr) evolution
of $\omega$ Cen based on 3D hydrodynamical evolution with feedback effects
of SNe II and SNe Ia on ICM. Although they reproduced the overall metallicity spread
observed in $\omega$ Cen, they failed to explain its $Y$ spread.
These previous models assumed that star formation can last more than
$\sim 1$ Gyr in the formation of $\omega$ Cen.

Recent observations have found, however, that there appears to be
no/little change in [Si/Fe]  and [Ca/Fe]  in stellar populations
of $\omega$ Cen (Johnson \& Pilachowski 2010, JP10; Magurno et al. 2019).
Since chemical enrichment by SNe Ia can decrease 
${\rm [\alpha/Fe]}$ with increasing [Fe/H],
the observed  lack of ${\rm [\alpha/Fe]}$ evolution
with [Fe/H] has been suggested to be no/little chemical enrichment
of ICM by SNe Ia in  $\omega$ Cen (JP10). 
These observations therefore imply that
star formation should be completed within an order of $10^8$ yr,
because 
some fraction of SNe Ia promptly enrich ICM within 1 Gyr (Totani et al. 2008).
Such a short formation timescale is inconsistent with
most of previous works for the ages of stellar populations in
$\omega$ Cen: see Table 1 for a summary of these works.
D'Antona et al. (2011) have investigated correlation/anti-correlation between
[O/Fe] and [Na/Fe] among stars with [Fe/H] ranging from $-2$ to $-0.3$ in
$\omega$ Cen (see their Fig. 1) and concluded that 
the formation timescale of $\omega$ Cen should be a few times $10^8$ yr: the
observed O and Na abundances matches well with yields from massive AGB 
stars (not from low-mass ones).
These recent results are inconsistent with the long star formation timescale
adopted in previous theoretical works. Thus, a new model needs to be constructed
that can self-consistently explain these  recent observations  as well as 
previous observations.

The purpose of this paper is thus to investigate chemical and dynamical evolution
of $\omega$ Cen in its early formation history based on a new model of $\omega$ Cen.
In the new model, $\omega$ Cen was formed from a  giant molecular cloud (GMC)
in the central region of its host dwarf galaxy that is 
a building block of the Galaxy.
Almost all stars of $\omega$ Cen
are formed from multiple episodes of star formation
within $\sim 300$ Myr 
so that ${\rm [\alpha/Fe]}$ of the stars ($\alpha$=Si, Ca, and Ti,
not O and Mg)  can be kept high 
(higher than the solar value).
Chemical enrichment in $\omega$ Cen  can proceed rapidly,
because SNe II from multi-generations of stars can pollute the  ICM
very efficiently.
Gas ejected from AGB stars and NSM can be mixed with ejecta from SNe II,
which ended up with the formation
of stars with high $Y$
and with large variations in $r$- and $s$-process elements.

The plan of the paper is as follows.
We describe the models of $\omega$ Cen formation
in  a dwarf galaxy 
in \S 2.
We present the results of chemical evolution of forming $\omega$ Cen 
in \S 3.
Based on these results,
we provide several implications of the results
and discuss how well the new model can reproduce the existing observations
in \S 4.
We summarize our  conclusions in \S 5.
Our previous works already discussed the formation of $\omega$ Cen from
a nucleated dwarf galaxy (BF03),
dynamical evolution of multiple cluster systems formed from GMCs
(Bekki 2017; B17a),
and GC formation  within
massive fractal GMCs (Bekki 2017b; B17b).
We therefore focus exclusively on the chemical properties of $\omega$
Cen and their spatial variations in the present study.
We only focus on  the chemical evolution of 12 elements 
(H, He, C, N, O, Na, Mg, Al, Fe, Ba, La, and Eu) and does not discuss extensively
other elements such as Pb and Mn in the present study
though these abundances have also fossil information
on the chemical evolution of $\omega$ Cen (e.g., Cunha et al. 2010;
D'Orazi et al. 2011, D11; Romano et al. 2011; Pancino et al. 2011).

\newpage
\begin{deluxetable}{ll}
\footnotesize  
\tablecaption{References for previous
works that suggested  long ($>1$ Gyr) 
and short ($<1$ Gyr) formation timescale of $\omega$ Cen.
See Stanford et al. (2006) for a detailed discussion on the methods
to derive ages of stellar populations of $\omega$ Cen in these works.
\label{tbl-1}}
\tablewidth{0pt} 
\tablehead{
}
\startdata
Long timescale  &  Norris \& Da Costa (1995), Hilker \& Richtler (2000),  \\
 &
Hughes \& Wallerstein (2000), Smith et al. (2000), Lee et al. (2002), \\
&
Origlia et al. (2003), Sollima et al. (2005), Stanford et al. (2006), \\
& Villanova et al. (2007)  \\
Short timescale  &  Pancino et al. (2002), Ferraro et al. (2004),  \\
 &
D'Antona et al. (2011), D'Orazi et al. (2011), this work \\
\enddata
\end{deluxetable}

\section{The model}

\subsection{Overview}

In the present formation scenario of $\omega$ Cen,
a metal-poor (${\rm [Fe/H] \sim -1.7}$) GMC with the total  mass
of $\sim 10^7 {\rm M}_{\odot}$ was first formed in the central
region of a gas-rich dwarf galaxy that was a building block of the Galaxy.
Accordingly, the GMC was initially surrounded by low-metallicity ([Fe/H]$<-1.7$) 
field stars, a fraction of which was later gravitationally bound by $\omega$ Cen to be
metal-poor stars with $-2.2 < {\rm [Fe/H]} < -1.7$ observed
in $\omega$ Cen (e.g., Rey et al. 2000; JP10;  Bono et al. 2019).
Although the formation process of such a very massive GMC is not specified 
in the present study,  tidal interaction of the dwarf with the Galaxy
could be responsible for the formation of such a GMC.
After the initial bursty formation of first generation
(1G)  stars within the GMC,
a large number of SNe II can expel the significant amount of gas left over
from the burst.  Gas ejected from
1G AGB ejecta  can be trapped by the deep potential
well of the host dwarf so that second generation
(2G) of stars can be formed from
the gas.  
The ejecta of  SNe II from  2G population can be trapped in
the central region of the dwarf and consequently
mixed with gas ejected from 1G AGB stars to form
third generation of  stars. 
External accretion of pristine gas from interstellar medium
(ISM)  onto $\omega$ Cen for mixing with AGB ejecta is not considered
in the present study,
though such accretion is a key process in the formation of other massive
GCs like Terzan 5 (e.g., McKenzie \& Bekki 2018) and 47 Tuc
(McKenzie \& Bekki 2019, in preparation).

Gas ejected from NSMs can be trapped by the ICM when a large amount of 
AGB ejecta
is accumulated in the central region of 1G stars,
because the ejecta from NSMs can lose energy and momentum through
interaction with the high-density  ICM (see BT17 for discussion on this point).
Accordingly,  AGB ejecta is crucial for the formation of new stars
from gas polluted by NSM in the present scenario.
Although NSM ejecta can significantly increase [Eu/Fe] in the ICM,
SNe II ejecta from later generations (LG) of stars can decrease [Eu/Fe].
Therefore,  NSM events in LG need to be incorporated in the present
chemodynamical model so that the observed flat [Eu/Fe]$-$[Fe/H] relation
can be reproduced. 
The formation of new stars from AGB and NSM ejecta
mixed with SNe II ejecta can continue over $\sim 300$ Myr until AGB ejecta
is expelled either by SNe Ia or by ram pressure stripping by the Galactic
warm/hot gas. During the destruction of the host dwarf, the initially
massive $\omega$ Cen can lose a significant fraction of its 1G stars 
to the early Galactic halo. We neither model the truncation of star formation
by SNe Ia or ram pressure nor investigate the destruction process of 
the host dwarf in the present study.

The present study does not discuss the origin of metal-poor stars
with ([Fe/H]$<-1.7$) observed in $\omega$ Cen (JP10) in a quantitative manner.
We consider a scenario
in which these metal-poor stars existed before the formation of the natal
GMC of $\omega$ Cen and were gravitationally trapped by $\omega$ Cen
after its formation.
Our previous simulations of nucleated dwarfs showed that a minor fraction
of field stars around stellar nuclei/nuclear star clusters can be still
gravitationally trapped by the nuclei/clusters even after the destruction
of their host dwarfs (Bekki \& Yong 2012).
Therefore, the above scenario can be regarded as  quite convincing and realistic.
We use our own original simulation code that can be run
on GPU clusters (Bekki 2013a)
in order to perform smooth particle hydrodynamics (SPH)
simulations
of GC formation within massive MCs.

\subsection{Fractal GMC}

Since the details of fractal GMC models adopted in the present study
are given in B17b,  we here briefly describe the models.
A GMC with a size $R_{\rm gmc}$ and a mass $M_{\rm gmc}$ is assumed
to be represented by $\sim 10^6$ SPH particles and have fractal
structures within it.
A GMC is assumed to have a power-law radial density profile
($\rho_{\rm mc}(r)$) as follows:
\begin{equation}
\rho_{\rm gmc} (r)=\frac{\rho_{gmc,0}}{ (r+c_{\rm gmc})^{\beta} },
\end{equation}
where $r$,  $\rho_{\rm gmc, 0}$,  and $c_{\rm gmc}$,
$\beta$  are the distance from the GMC's
center,  a constant that is determined by $M_{\rm gmc}$ and $R_{\rm gmc}$,
the core radius of the GMC, and the power-law slope, which 
is fixed at $1$ in the present study.
The GMC is assumed to have a fractal gaseous distribution  characterized by
a fractal dimension $D_3$ that is fixed at 2 in
the present study.

The initial virial ratio ($t_{\rm vir}$) 
of a GMC  is described as follows:
\begin{equation}
t_{\rm vir}=\frac{ 2T_{\rm kin} }{ | W_{\rm gmc} | },
\end{equation}
where $W_{\rm gmc}$ is the initial total potential energy of the GMC
and $T_{\rm kin}$ is the total kinetic energy due to random motion
and global rotation in the GMC.
We present only the results of the models with $t_{\rm vir}=0.4$,
We also  consider
rigid rotation of a GMC in some models, because previous observations suggested
that velocity gradients within MCs could be due to
such rotation (e.g., Phillips 1999; Rosolowsky et al. 2003). 
Accordingly,
$T_{\rm kin}$ is the combination of the total random energy
$T_{\rm ran}$ and the total rotational one ($T_{\rm rot}$),
and the ratio of the two 
($f_{\rm rot}$) is a parameter in the present study as follows:
\begin{equation}
f_{\rm rot}=\frac{ T_{\rm rot} }{ T_{\rm kin} } .
\end{equation}
We mainly show the results of the models with $f_{\rm rot}=0.01$ 
in the present study, because recent observations show that the angular
momentum of GMCs is quite small (e.g., Rosolowsky et al. 2003).
We also investigate the models with different $f_{\rm rot}$ to understand
how $f_{\rm rot}$ changes the structure and kinematics of $\omega$ Cen.
Initial gaseous temperature and metallicity are
set to be 10K and [Fe/H]=$-1.7$ in all MCs.
The early chemical enrichment and feedback effects in natal
GMCs  by stellar winds of OB stars
(e.g., Bekki \& Chiba 2007) are not included in the present study.

\subsection{Host dwarf}

The most important role of $\omega$ Cen's host dwarf is that its deep
gravitational potential can retain gas ejected from stars during
the intense star formation.
Since the host dwarf should be dominated by dark matter,
we only consider the mass distribution of dark matter.
We adopt the density distribution of the NFW
halo (Navarro, Frenk \& White 1996) derived  from $\Lambda$CDM simulations
in order to describe the radial mass density profile of a dark matter halo
in a disk galaxy as follows:
\begin{equation}
{\rho}(r)=\frac{\rho_{0}}{(r/r_{\rm s})(1+r/r_{\rm s})^2},
\end{equation}
where  $r$, $\rho_{0}$, and $r_{\rm s}$ are
the spherical radius,  the characteristic  density of a dark halo,  and the
scale
length of the halo, respectively.
The $c$-parameter ($c=r_{\rm vir}/r_{\rm s}$, where $r_{\rm vir}$ is the virial
radius of a dark matter halo) and $r_{\rm vir}$ are chosen appropriately
as 16 for the low-mass dwarf with the total mass of $10^{10} M_{\odot}$
and $r_{\rm s}=1.2$ kpc.

\subsection{Star formation}

We consider that if the mass density of a gas particle ($\rho_{\rm g}$) is
higher than the threshold gas density for star formation ($\rho_{\rm g, th}$),
then the gas particle is converted into a new stellar particle. Therefore,
the physical condition for star formation is as follows:
\begin{equation}
\rho_{\rm g} > \rho_{\rm g, th}.
\end{equation}
Since the typical mass density of the core of a GMC is $10^5$ atom cm$^{-3}$
(e.g.,  Bergin \& Tafalla 2007),
we adopt $\rho_{\rm g, th}=10^5$ atom cm$^{-3}$ as a reasonable value.
In the present study, we consider that the IMF can be different between
new stars formed from original gas of a GC-forming GMC (1G stars)  and those formed 
from gas ejected from 1G stars (2G stars). This is mainly because our
recent simulations have demonstrated the IMF in 2G star formation
 to be more top-light (i.e., a smaller number of SNe II progenitors)  owing
to dynamical influences of existing dense stellar systems on gas from 1G stars
(Bekki 2019a, B19a).
1G stars are assumed to have  the canonical  Salpeter IMF with the slope
of $-2.35$ and the lower-mass and upper-mass cutoffs being 
$0.1 M_{\odot}$ and $50 M_{\odot}$, respectively.
This might not be a reasonable choice, 
because our previous simulations showed that the IMF can be top-heavy
in the nuclear starburst regions of galaxies
(Bekki 2013b), which could be relevant
to the formation cite of $\omega$ Cen in the present study.
However, we use the canonical IMF and discuss the problem of this IMF
in reproducing the observed properties of $\omega$ Cen in \S 4.

2G stars are assumed to  have the top-light IMF with the slope of $-2.35$ and
the mass fraction of SNe II ($f_{\rm sn2}$)
for 2G stars being 0.02
in all models of the present study. This $f_{\rm sn2}$ is by a factor of 7
smaller than that for 1G stars.
We have investigated the models with different $f_{\rm sn2}$ in order to
find the best model for $\omega$ Cen in which the maximum [Fe/H] in the stars
can be as large as $-0.3$. We have found that $f_{\rm sn2}$ should be 
as low as $0.02$, because the models with larger $f_{\rm sn2}$ shows
the maximum [Fe/H] larger than 0. Also, the adopted models with the top-light
IMF means weaker SNe II feedback effects thus a larger amount of ICM being
converted into new stars. Thus, we describe the results mainly for the models
with $f_{\rm sn2}=0.02$ in the present study.

\newpage
\begin{deluxetable}{lllll}
\footnotesize  
\tablecaption{The adopted [La/Fe] for different stellar masses
predicted from previous works (F14).
Four different models for the yields are investigated for comparison
and
the yields for Y1 model are exactly the same as those in F14.
The La yields for Y2, Y3, and Y4 are increased by 0.2, 0.25, and 0.3 dex,
respectively, with respect to Y1.
The Ba yields in Y2, Y3, and Y4 are also increased in the same way
as La  accordingly.
\label{tbl-1}}
\tablewidth{0pt} 
\tablehead{
}
\startdata
Model ID  &  
$4M_{\odot}$  & $5M_{\odot}$  & $6M_{\odot}$  & $7M_{\odot}$  \\ 
Y1 &  0.72 & 0.64 & 0.52 & 0.20 \\
Y2 &  0.92 & 0.84 & 0.72 & 0.40 \\
Y3 &  0.97 & 0.89 & 0.77 & 0.45 \\
Y4 &  1.02 & 0.94 & 0.82 & 0.50 \\
\enddata
\end{deluxetable}

\subsection{Chemical enrichment}

\subsection{Chemical yields}

In the present study, we investigate the time evolution of chemical
abundances of 12 elements, H, He, C, N, O, Na, Mg, Al, Fe, La, Ba, and Eu
in each model. Chemical enrichment in forming $\omega$ Cen is due to
gas ejected from SNe II, AGB stars, and NSM: ejecta from SNe Ia is not 
considered, because chemical enrichment by SNe Ia ends up with a significant
decrease of ${\rm [\alpha/Fe]}$, which is not observed in JP10. In order
to model chemical evolution of $\omega$ Cen, we adopt chemical yields
from Kobayashi et al. (2006, K06) 
for SNe II, Fishlock et al. (2014; F14) for AGB stars,
and Tsujimoto \& Shigeyama (2014) for NSMs.
Chemical yield tables appropriate for low-metallicity stars
are chosen from these theoretical predictions by K06 and F14.
It should be noted here that chemical yields  used in B17b
are different from F14 adopted in the present study.

\subsubsection{SNe II}

Since chemical enrichment processes 
by SNe II are described by B17b in detail too,
we briefly summarize them here.
Gaseous ejecta from  a SNe II can mix with its surrounding SPH particles
so that the gas particles can increase their chemical abundances.
Each SN can eject SPH gas particle  with an initial ejection
speed of $v_{\rm ej}$, which is estimated from the following equation:
\begin{equation}
f_{\rm kin}E_{\rm sn}=0.5 (m_{\rm s}-m_{\rm rem})v_{\rm ej}^2,
\end{equation}
where $f_{\rm kin}$, which is set to be 1, is the fraction of kinetic energy
in total SN energy and 
$m_{\rm rem}$ is the mass of a neutron star or a black hole
that is left after SN explosion
for massive stars. 
Neutron stars that are left after SNe II can be the source of Eu in
the present study (if they are pairs of neutron stars). 
For $m_{\rm s}=8 {\rm M}_{\odot}$ and $f_{\rm kin}=1$,
$v_{\rm ej}=3800$ km s$^{-1}$ ($m_{\rm ej}=6.5 {\rm M}_{\odot}$) and 
we adopt these values for all SNe II.
The kinetic energy of a SN is distributed equally among SPH gas particles
surrounding the SN. 

\subsubsection{SNe Ia}

We consider that chemical enrichment by SNe Ia should not proceed
in the early formation phase of $\omega$ Cen 
owing to no/little evolution
of ${\rm [\alpha/Fe]}$  with [Fe/H] (JP10).
Accordingly, chemical enrichment by SNe Ia is not included
in all models of the present study.
This assumption of no SNe Ia over the period of
$\sim 300$ Myr in GC formation would not be so reasonable,
if SNe Ia can occur only  $\sim 10^8$ yr  after star formation
(``prompt SNe Ia''). However, it is observationally difficult
to constrain the minimum delay time of SNe Ia 
(i.e., time lag between the formation of a binary pair of
stars and the onset of SNe Ia explosion from the binary merging)
through a comparison between chemical evolution models and observations
(e.g., see  Figure 9 in Siegel et al. 2019). 
Therefore, the assumption of no SNe Ia within a few Myr is not unreasonable in 
the present study.
It would be possible that all ICM can be removed
by other physical processes (such as ram pressure stripping) $\sim 300$
Myr after star formation in $\omega$ Cen. If this is the case,
SNe Ia ejecta cannot mix with ICM to form new stars:
SNe Ia cannot contribute to  the chemical evolution of $\omega$ Cen.
We will discuss this point later in \S 4.

\subsubsection{AGB}

Each AGB star is assumed to eject gas with a wind velocity of 
$\sim 10$ km s$^{-1}$ and the chemical abundances consistent with
the adopted chemical yield table (F14). In order to model star formation
directly from AGB ejecta without being diluted by gas, we adopt the following
``AGB particle'' model (Bekki 2019b; B19b).
In the AGB particle model, soon after a new star enters into its AGB phase,
a new SPH gas particle (``AGB particle'')
is ejected from the star with its initial speed  ($v_{\rm w}$)
of 10 km s$^{-1}$ with respect to the star. Accordingly,
if this new gas particle is converted into a new star,
then the new star can have chemical abundances that are the same as the AGB 
ejecta from which the gas originates. 
In previous galaxy-scale simulations (B13, B15),
AGB ejecta is mixed with neighboring gas particles so that the ejecta
can be diluted by ISM to have chemical abundances similar to those of ISM.
The new AGB particle method can avoid this dilution that does not always occur
in ICM.
The new AGB particle can be mixed
with SNe II and NSM ejecta if it is close to SNe II and NSM events. 

There is a significant uncertainty in the predicted AGB  yields for [Ba/Fe]
and [La/Fe] over $-1.7 < {\rm [Fe/H]} < -0.3$ 
(e.g., Busso et al. 2001; see also Figure 1 in Tsujimoto \& Bekki 2012,
which shows $\sim 2$ orders of magnitudes difference in the predicted
yields for $-2  <{\rm [Fe/H]} < -1$).
Furthermore, the low yields for  La predicted in F14
(e.g., [Ba/Fe]$\sim 0.4$ 
and [La/Fe]$\sim 0.2$
for $m=7 M_{\odot}$) 
cannot simply explain the observed significant fraction of
stars with [La/Fe]$\sim 0.5$ in $\omega$ Cen (e.g., JP10).
We therefore investigate different sets of models in which [Ba/Fe] and [La/Fe]
are different by a factor of 3 in order to find the best set of models that
can reproduce the observed range of [Ba/Fe] and [La/Fe]. We mainly
shows the models in which Ba and La yield are by a factor of 2 larger than
those shown in F14, because the models show that [Ba/Fe] and [La/Fe] at
higher metallicities ($>-1$)  are more consistent with observations.
The four yield models for AGB stars used 
in the present study (Y1, Y2, Y3, and Y4 with Y1 being exactly
the same as F14)  are briefly summarized in Table 2.
We also compare between the results from different models.

AGB stars from 2G population can also chemically enrich ICM in the later
phase of $\omega$ Cen formation, though such enrichment processes
are much less significant than those by 1G AGB stars.
However, it is possible that the time evolution of $s$-process elements
with [Fe/H] can be significantly influenced by the 2G AGB stars,
because [Ba/Fe] and [La/Fe] of AGB ejecta depends strongly on
[Fe/H] (see Figure 1 in Tsujimoto \& Bekki 2012, TB12).
We therefore incorporate such chemical enrichment processes into some models based
on the following relation (TB12) for [Fe/H]$<-1$:
\begin{equation}
{\rm [Ba/Fe] = [Ba/Fe]_0+ 0.6 ([Fe/H]-[Fe/H]_0}),
\end{equation}
where [Ba/Fe]$_0$ is [Ba/Fe] at [Fe/H]$_0=-1.7$.
Once a 2G star enters into the AGB phase, the neighboring gas particles are 
assumed to be chemically polluted by the AGB wind. The Ba and La yields
are calculated from (i) F14 and (ii) the above equation for [Fe/H] of the star, and the 
chemical abundances of the gas particles are changed accordingly.
Although we investigated how the existing field AGB stars of $\omega$ Cen's
host dwarf galaxy influence the chemical evolution of $s$-process elements,
we found that  such chemical enrichment processes by the field stars
do not change the present results as long as
the mass ratio
of the field stars to the original GMC ($m_{2}$) is less than 3.
Therefore, we do not show these results in the present study.

\subsubsection{NSM}

The details of mixing processes of NSM in forming GCs
are yet to be fully understood,
because no hydrodynamical simulations on this have ever been done.
All of the NSM ejecta is assumed to be mixed well with and consequently
retained in ISM of galaxies
in our recent study 
of galactic chemical evolution (e.g., Tsujimoto et al. 2017).
The assumed 100\% retention  of NSM ejecta in ISM would not be 
so realistic for ICM, because the total gas mass of ICM is much
smaller than ISM of galaxies (BT17).
Accordingly, we here adopt two different mixing models: (i) density-dependent
and (ii) uniform mixing models.
The density-dependent model is described as follows.
BT17 showed that  gaseous ejecta from NSM can be trapped by ICM within a GC,
only
if the density of ICM ($\rho_{\rm icm}$)
is rather high ($>10^4$ atom cm$^{-3}$). In order to model this retention 
of NSM ejecta by ICM, we use the following formula for
the stopping length ($l_{\rm s}$)  of NSM ejecta:
\begin{equation}
l_{\rm s}= l_{\rm s, 0} (\frac{ \rho_{icm} }{ \rho_{\rm icm, 0} })^{-1} {\rm pc},
\end{equation}
where $\rho_{\rm icm,0}$ is the reference density of the ICM that is set to be
$10^5$ atom cm$^{-3}$ (BT17) and
$l_{\rm s, 0}$ is the stropping length for $\rho_{\rm icm,0}$.
In this formula,
high-speed NSM ejecta can be retained by ICM only after it travels $l_{\rm s}$.
In the present study, we investigate the models
with $l_{\rm s, 0}=0.27$pc, 0.78 pc, and 2.3 pc.

The neighboring gas particles around one NSM event are first searched within 
$r \le r_{\rm nsm}$ of the NSM, where $r_{\rm nsm}$ is the mixing length of 
NSM ejecta.
Since the distance between a NS merger and one of its ($j$th) neighboring gas particles
($r_j$) can be larger than $l_{\rm s}$,
only a fraction of the total mass of the NSM ejecta can be retained
by ICM. The total mass of NSM ejecta retained by $j$th gas particle
($m_{\rm nsm, \it j}$) is described as follows:
\begin{equation}
m_{\rm nsm, \it j}= f_{\rm ret} \frac{ m_{\rm nsm} }{ N_{\rm nei} },
\end{equation}
where $m_{\rm nsm}$ is the total mass of NSM ejecta,
$f_{\rm ret}$ is the mass fraction of NSM ejecta retained by the gas particle,
and $N_{\rm nei}$ is the total number of neighboring gas particles around 
the NS merger. Here it is assumed that
NSM ejecta can be distributed among all of its neighboring
gas particles. 
In order to estimate $f_{\rm ret}$, 
we consider that the gas particle  has a spherical shape with 
a radius $h_{\rm j}$, where $h_{\rm j}$ is the SPH smoothing length,
and (ii) the flux per unit area for NSM ejecta at the position
of the gas particle is as follows:
\begin{equation}
F_{\rm nsm, \it j} = \frac{ m_{\rm nsm} }{ 4 \pi r_{\rm j}^2 } .
\end{equation}
Accordingly, $f_{\rm ret}$ is as follows:
\begin{equation}
f_{\rm ret} = \pi h_j^2 F_{\rm nsm, \it j} .
\end{equation}
If $r_j$ is much smaller than $h_j$, $f_{\rm ret}$ can be larger 
than 1 in the above formula. For that case, $f_{\rm ret}$ is set to be 1.

The uniform mixing model is as follows. All gas particles around each
NSM event can be uniformly polluted by NSM in this model,
even if the physical properties of the gas particles are quite different.
Accordingly, $f_{\rm ret}$ is a free parameter ranging from 0 to 1
and needs to be chosen such that the observed [Eu/Fe] of $\omega$ Cen
can be reproduced. Since $l_{\rm s}$ is not introduced in this model,
the mixing radius ($r_{\rm nsm}$) is also a free parameter.
We investigate  the models with $r_{\rm nsm}=\epsilon_{\rm g}$,
and $10 \epsilon_{\rm g}$, where 
$\epsilon_{\rm g}$ is the gravitational softening length of gas
particles. We mainly show the results of the density-dependent
models for mixing of NSM ejecta, because they are more realistic.

In order to calculate the number of NSM events  per unit stellar mass for a given
IMF, we assume that the ratio of NSM events per one SNe II ($r_{\rm nsm}$)
is $10^{-3}$. We however investigate models with different $r_{\rm nsm}$
in the present study.
The delay time distribution (DTD) of NSM ($N_{\rm nsm}(t)$) is as follows:
\begin{equation}
N_{\rm nsm} (t) = N_{\rm nsm, 0} (t/t_{\rm nsm})^{-1},
\end{equation}
where $N_{\rm nsm, 0}$ is the normalization factor and $t_{\rm nsm}$
is a parameter that determines the DTD (e.g., Dominik et al. 2012).
For most models, we assumed that NSM occurs between $10^7$ yr to $10^{10}$ yr
and $t_{\rm nsm}$ is set to be $10^7$ yr.
We also investigate the models with a different $N_{\rm nsm}(t)$ just for
comparison.
We consider that (i) $0.01 M_{\odot}$ can be ejected from one NSM event,
(ii) the mass fraction of Eu in the ejecta ($f_{\rm eu}$) is 0.01,
and (iii) [La/Eu] and [Ba/Eu] are set to be $-0.2$ and $-0.1$
, respectively, which are reasonable for stars with [Fe/H]$\sim -1.7$  $\omega$ Cen 
(e.g., Fig. 4 in D11).
If we consider  NSM ejecta consisting of elements with mass numbers
larger than 90, then  $f_{\rm eu}=4.8 \times 10^{-3}$ rather than
0.01. This factor of $\sim 2$ difference in Eu yield does not change
the present results so much.   

\begin{deluxetable}{ll}
\footnotesize  
\tablecaption{ Description of the parameter values
for the fiducial model. One of key elements in the model
is that new gas particles are ejected from one AGB star
(``AGB particle'' method) to allow the AGB ejecta to be
converted into new stars without dilution.
\label{tbl-1}}
\tablewidth{-2pt}
\tablehead{
}
\startdata
GMC mass & $M_{\rm gmc}=2 \times 10^7 M_{\odot}$  \\
GMC radius & $R_{\rm gmc}=140$ pc  \\
GMC scale radius & $R_{\rm gmc}=27$ pc  \\
GMC initial position  & $r_{\rm gmc}=0$ pc  \\
Initial virial ratio   & $t_{\rm vir}=0.4$  \\
Fraction of rotational energy   & $f_{\rm rot}=0.01$  \\
Fractal dimension   & $D_3=2$  \\
Mixing of NSM ejecta   & Density-dependent ($r_{\rm nsm}=0.55$ pc) \\
Threshold density for star formation  & $\rho_{\rm th}=10^5$ atom cm$^{-3}$ \\
Simulation code  & B17b  \\
Mass resolution  & $19 M_{\odot}$  \\
Size resolution  & $0.55$ pc  \\
AGB yield  &  F14  \\
SNe II yield  & K06   \\
NSM yield  & TS14   \\
SNe II feedback effects  &  Included   \\
SNe Ia feedback effects  &  Not included   \\
AGB feedback effects  &  Included   \\
AGB wind velocity  &  $10$ km s$^{-1}$   \\
AGB particle method  &   Included   \\
\enddata
\end{deluxetable}

\subsection{Parameter study}

We mainly describe the results for
the fiducial model  with $M_{\rm gmc}=2 \times 10^7 M_{\odot}$,
$R_{\rm gmc}=140$ pc,  $r_{\rm gmc}=0$ kpc (i.e., located in the center of
its host dwarf), firstly because the final stellar mass of the simulated GC
can be as large as $\sim 10^7 M_{\odot}$ that is $2-3$ times 
larger than the present-day mass ($\sim 4 \times 10^6 M_{\odot}$),
and secondly because the total mass of He-rich stars can be as
large as $[6-8]\times 10^5 M_{\odot}$ as observed (e.g., Norris 2004).
The initial total number of particles ($N$) is set to be 1040000 in the fiducial model,
and $N$ increases significantly owing to the ejection of new AGB particles from
AGB stars 
during a simulation.
The mass and size resolutions are $10 M_{\odot}$ and 2 pc, respectively, in the 
fiducial model. The values of basic parameters adopted in the fiducial model
are given in Table 3.

We investigate how the total masses, sizes, and initial locations of GMCs
determine the final physical properties of GCs by changing the basic parameters,
$M_{\rm gmc}$, $R_{\rm gmc}$, and $r_{\rm gmc}$. We also investigate
how the present results depend on the model parameters for mixing processes
of NSM ejecta (e.g., uniform or density dependent mixing)
in order to make robust conclusion on the results.
In the present study, [Ba/Fe]$-$[Fe/H], [La/Fe]$-$[Fe/H],
and  [Eu/Fe]$-$[Fe/H] relations depend strongly on
the details of the modeling for NSM ejecta mixing.
We therefore extensively investigate  how these relations can be controlled
by model parameters for NSM ejecta mixing.
We show the results of only 22 representative models, though we have
investigated 30+ models.
The parameter values of
these representative models with are summarized in Table 4.
In the followings,  1G and 2G  stars means 
(i) new stars formed from original gas particles
and those formed from gas ejected from AGB stars, whenever they are formed.
Accordingly, new stars formed from re-accreted original gas after 2G star formation
has started are also identified as ``1G'', though they can be younger than some
of the 2G stars formed earlier.

\section{Results}

\subsection{The fiducial model}

\subsubsection{Spatial distributions and kinematics
  of He-poor and He-rich stars}

Figure 1 shows how the projected distributions of ``He-poor''($Y \le 0.25$)
and ``He-rich'' ($Y>0.25$) stars at $T=300$ Myr  are different 
in the simulated 
$\omega$ Cen for the fiducial model M1. Clearly, both He-poor and He-rich
populations
shows rather flattened shapes in all three projections with the major
axes of the shapes of the two populations being aligned with each other.
The He-rich population shows a stronger central mass concentration, which
reflects that it is formed from gas  that
is ejected from AGB stars  and then accumulated into the deeper potential
well of the forming GC. Although the simulated flattened distributions
are qualitatively consistent with the observed shape of $\omega$ Cen,
it should be noted here that the present simulations does not investigate
long-term two-body relaxation processes that significantly changes
the shapes of GCs (e.g., Mastrobuono-Battisti \& Peters 2016).
It could be possible that the simulated
GC can become significantly less flattened owing to such dynamical
effects, though the two body relaxation timescale is longer for
$\omega$ Cen.

Figure 2 demonstrates that the He-poor population has a small amount of
rotation ($V_{\rm max} \sim 2$ km s$^{-1}$, where $V_{\rm max}$ is
the maximum rotational speed) within the central 30 pc in the three
2D kinematic maps.  This global
rotation of the He-poor population inherits from the initial angular momentum
of its natal GMC with $f_{\rm rot}=0.01$. Accordingly, the final amplitude
of the rotation in the He-poor population of
the simulated $\omega$ Cen  depends strongly on $f_{\rm rot}$
in the present study.
The He-rich population, on the other hand, does not show clear rotation
in the three 2D kinematic maps. This result is in a striking contrast
with our previous works (Bekki 2011), which demonstrates that 2G stars
formed from 1G AGB ejecta  in a GC
(corresponding to the He-rich population in the present study)
can have a significant amplitude of rotation within the GC. 

Significant differences between the present study and our previous ones
are as follows. First, the present study simulates GC formation from
initial star formation from natal GMCs 
(i.e., 1G formation) of GCs whereas previous ones investigated
star formation from AGB ejecta in existing (1G) stellar systems. Second,
the feedback effects of SNe II from 2G stars are included in the present
study, but not in previous ones. Third, all AGB ejecta is assumed to be 
similarly rotating with respect to the center of the existing stellar system 
in previous works, whereas such initial uniform rotation is not assumed
in the present study: the initial angular momentum of AGB ejecta can be
quite different between different AGB stars.
The second difference is the main physical reason for the less amount
of rotation in the He-rich stars in the fiducial model. 
SNe II feedback effects can stir the ICM so that the new stars formed from
the ICM cannot have a large degree of ordered motion in the present study.

\subsubsection{Abundance distribution functions (ADFs)}

Figure 3 describes the abundance distribution
functions (ADFs)  of the 11 elements investigated in the fiducial 
model. The strong peak in each abundance corresponds to the abundance
of 1G population and thus to that of its natal GMC. Unlike the observed
ADF of [Fe/H] in $\omega$ Cen, the simulated ADF does not show clearly
distinct multiple peaks around [Fe/H]$\sim -1.4$ (corresponding to
``RGB-Int1''; see Figure 8 in JP10),
$\sim -1.1$ (``RGB-Int2+3''), and $\-0.6$ (``RGB-a'').
Also, the observed ADF shows a broad peak around [Fe/H]$\sim -1.7$
(``RGB-MP''), which is not {\it assumed} in the present model:
the simulated very sharp peak around [Fe/H]$\sim -1.7$ is due simply
to the adopted assumption of a single metallicity in the GMC.
If a GC-forming GMC is  initially very massive,
as assumed for $\omega$ Cen in the present work,
the GMC can possibly have a significant metallicity spread,
which can end up with a broad peak of the ADF for 1G stars formed within
the GMC.

The  ADF of $Y$ in this model shows three peaks around [Fe/H]$\approx 0.25$,
$\approx 0.31$, and $\approx 0.36$. In the adopted chemical yields for AGB stars
(F14), massive AGB stars can eject gas with higher $Y$ ($\sim 0.36$).
Therefore, the third peak around $Y\sim 0.36$ corresponds to the 
secondary star formation from gas ejected from massive AGB stars 
($m>6 {\rm M}_{\odot}$).
The second peak in the ADF of $Y$ ($\sim 0.31$) is due to the formation
of new stars from gas ejected from less massive AGB stars.
As shown in Figure 4, the third peak of $Y$ is the most
pronounced for stars with $-1.6 < {\rm [Fe/H]} \le -1.3$, which means 
that a larger fraction of new stars with high $Y$ can be formed
AGB ejecta that is not so much polluted by SNe II ejecta.
Clearly, there are stars with normal $Y$ ($\sim 0.25$) yet
higher [Fe/H] ($>-1.6$) in Figure 4, though the number fraction
of such stars is quite small.   These He-poor and Fe-rich stars
are formed from pristine gas that is later accreted onto the existing
stellar system, partially polluted by SNe II ejecta, 
and finally converted into new stars
in the central region of the system.

Figure 3 demonstrates that the simulated ADF is bimodal for
[C/Fe], [N/Fe], [O/Fe], and [Na/Fe]. The physical origin of
these bimodal ADFs is caused by the efficient formation of 2G
new stars almost directly from AGB ejecta with low [O/Fe] and high [C/Fe],
[N/Fe], and [Na/Fe]. 
In this model, no dilution of AGB ejecta by pristine (original) gas can
occur so that the chemical abundance patters of some 2G stars can be 
almost exactly the same as  those of AGB ejecta.
This no/little  dilution of AGB ejecta is one of characteristics of $\omega$ Cen's
chemical evolution: it should be noted here that dilution of AGB ejecta
by pristine gas is essential for explaining chemical abundance patters of
other ``ordinary'' GCs with multiple stellar populations (e.g., B07).
As shown in Figure 4,
the locations
the peaks in the [O/Fe] distribution is significantly
different for the two [Fe/H] ranges, i.e., [Fe/H]$\le -1.6$ and
$-1.6<$[Fe/H]$\le -1.3$. However,
the peak locations of the distributions
in the three ranges for $-1.6<$[Fe/H] are not so different.
These [Fe/H]-dependent ADFs can be seen for [Na/Fe] in Figure 4,
though the ADF of [Na/Fe] for higher [Fe/H] is significantly
wider than that of [O/Fe]. It should be stressed that there
are stars with rather high [Na/Fe] ($>1$) only for the highest
[Fe/H] range ($>-0.5$), though the number fraction is quite small.
These high-[Na/Fe] stars can be formed from gas polluted by
SNe II whose progenitor stars have higher metallicities in
the chemical yield table by K06.

As shown in Figure 3,
the ADFs of [Mg/Fe] and [Al/Fe] are not bimodal
and they are not so broadened by the formation
of 2G stars
from AGB ejecta.
This is mainly because [Mg/Fe] and [Al/Fe] in the adopted 
AGB yields (for 2G stars) are not so much different from those adopted for
the initial GMC.
The ADFs of the two $s$-process elements,  [Ba/Fe] and [La/Fe],
have a strong peak around the initial abundances
with a very wide ``skirt'' ranging from $-1$ to 1.
Since Ba and La can be produced both by AGB stars and NSM
at different epochs,
their ADFs can be significantly broadened  after the formation of 2G stars
from AGB ejecta mixed with NSM one.
The ADF of [Ba/Fe] for [Ba/Fe]$>0$
has a broad peak around
[Ba/Fe]$\approx 0.7-1$ in this model.

As shown in Figure 4, 
the location(s) of the peak(s) and the level of broadening
in the ADF of [La/Fe] are quite different between different [Fe/H] ranges.
These simulated ADFs of [La/Fe] for different metallicity ranges
are not so consistent with the observed ones by JP10 (see Figure 12 of JP10).
For example, the observed ADF does not show its peak at [La/Fe]$<0$,
and there is no stars with [La/Fe]$<0$ for [Fe/H]$>-0.9$ (JP10).
The simulated $\omega$ Cen, however, has 
a large fraction of stars with 
[La/Fe]$<0$ for [Fe/H]$>-0.9$ and the peak around [La/Fe]$\sim 0.1$
for [Fe/H]$>-0.9$.
Such a large fraction of stars with low [La/Fe]
is due to star formation from
gas polluted too much by SNe II,
because SNe II ejecta  can lower [La/Fe] of ICM.
The observed ADF of [La/Fe] has only one peak 
for $-1.6 < {\rm [Fe/H] } < -1.3$
and for $-1.3 < {\rm [Fe/H] } < -0.9$
whereas the simulated one show two peak for the two [Fe/H] ranges.
These results that are not so consistent with observations
suggest that the present chemodynamical study
misses key physical processes
in the mixing of AGB and SNe II ejecta.

Observations showed that [Eu/Fe] ranges from $\sim -0.5$ to $\sim 0.8$
in $\omega$ Cen (Figure 17 of JP10), which is a significantly wider
distribution compared to the Galactic halo stars with metallicities similar
to $\omega$ Cen. There are a 
fraction of stars that have negative [Eu/Fe] in $\omega$ Cen, though there are no such
stars in the Galactic halo with the same metallicity
range as $\omega$ Cen (Figure 17 of JP10). Such stars with negative
[Eu/Fe] for [Fe/H]$>-1.7$ are indeed formed in the fiducial model, as shown
in Figure 3. The physical reason for this is as follows.
Although NSMs can increase [Eu/Fe] of ICM and thus [Eu/Fe] of
new stars,
such an increase cannot be so dramatic due to the retention capability of
NSM ejecta by ICM.
First of all, the total mass of ICM during 2G formation cannot be 
as large as $\sim 10^6 M_{\odot}$ that is required for the full mixing
of NSM ejecta in ISM of galaxies (Tsujimoto et al. 2017), 
firstly because the total mass
of AGB ejecta (i.e., the dominant component of ICM)
can be at most 5\% of 1G stars,
secondly because the AGB ejecta continues to  be consumed by
star formation.
Also,  the mass density of ICM around some  NSM events can be quite low 
so that NSM ejecta cannot be retained so well: rather high mass-density is required
for efficient retention of NSM ejecta
in this model with the density-dependent mixing model.
Thus, NSM ejecta can be only partially retained in ICM of forming $\omega$ Cen
in this model.

Owing to the partial retention of NSM ejecta,
[Eu/Fe] decrease of ICM due to chemical enrichment by SNe II 
can be  more
significant than [Eu/Fe] increase by NSMs in some local regions
of ICM.
Accordingly,  new stars formed from such local regions
can have negative  [Eu/Fe]: those formed from local regions where SNe II chemical 
enrichment is much less efficient can still have positive [Eu/Fe].
Therefore, [Eu/Fe] of a 2G star depends on whether the natal gas of the star
is polluted by NSM or SNe II to a greater  extent.
If all NSM ejecta can be retained in ICM, then
[Eu/Fe] can be as large as $\sim 2$, as shown later. The observed lack of such stars
with large [Eu/Fe] in  $\omega$ Cen means that only a small portion
of NSM ejecta could be retained by ICM
in the early formation phase of $\omega$ Cen.
Stars with low [Eu/Fe] ($<0$) at [Fe/H]$>-1.7$ observed in $\omega$ Cen
do not exist  in the Galactic stellar halo (JP10),
which implies  that the chemical enrichment process of $r$-process elements can be quite
different between $\omega$ Cen and the building blocks of the Galactic stellar halo.

\subsubsection{Radial abundance gradients}
As shown in Figure 5,  the simulated $\omega$ Cen has a negative radial
gradient of $Y$ within the central 10pc,
which is also consistent with
the more compact distribution of He-rich stars in Figure 1.
This strong concentration of 2G stars with high $Y$ (and low [O/Fe])
is also consistent with observations by Sollima et al. (2007)
and Gratton et al. (2011).
The main reason for this negative gradient is that
He-rich stars from AGB ejecta can be formed 
mostly in the central region of this GC where the low velocity AGB wind
can be captured efficiently.  The $1\sigma$ dispersion
in $Y$ at each radial bin
is quite large ($\sim 0.05$), which is due largely
to the co-existence of stars with rather high $Y$ ($\sim 0.34$) and
normal $Y$ ($\sim 0.25$) in the central 10pc.

In order to compare these results with corresponding observations, we here use
the observed number ratio ($f_{\rm bms}$)
of blue main-sequence (bMS) to red main-sequence
stars (rMS) in $\omega$ Cen (Sollima et al. 2007).  The number ratio
as a function of $R$ (distance from $\omega$ Cen's center in units
of arcmin) is described
as follows:
\begin{equation}
f_{\rm bms}=0.16 - 0.027 (R - 12).
\end{equation}
By assuming that $Y$ is 0.25 for rMS and 0.35 for bMS, we can convert
this relation into a radial dependence of $Y$. 
The blue dotted line in
Figure 5 describes the  negative slope of $Y$ observed in $\omega$
Cen, though the observed $f_{\rm bms}$ shows an apparently sharp decline with
increasing $R$.
 Only 0.01 change in $Y$ over 10 pc in the observed
$Y$ gradient of $\omega$ Cen is significantly shallower
than the simulated one. The simulated steeper $Y$ gradient at the birth
of $\omega$ Cen  might not be 
a serious problem, because the long-term dynamical evolution can make the gradient
shallower.

Clearly, the simulated GC shows a negative radial gradient of [Fe/H]
too, which is not so consistent with the observed apparently
flat  gradient in JP10, however.
The simulated [Fe/H] gradient  means that a significant fraction of new stars can be 
formed from AGB ejecta in the central region
of $\omega$ Cen after the ejecta is mixed well with
SNe II ejecta. The large $1\sigma$ dispersion in each [Fe/H] radial bin
is caused by the co-existence of stars formed from ICM with different
[Fe/H] at different epochs.
Since population mixing due to the long-term dynamical evolution
of the cluster caused by two-body relaxation effects can possibly
smooth out the original radial gradient of [Fe/H],
the radial gradient can possibly become flatter as observed.
However, we need to investigate
this point in our future work.

Although the radial gradients of [N/Fe] and [C/Fe] are weak and negative 
($\sim -0.06$ dex pc$^{-1}$) in the simulated $\omega$ Cen,
the radial [O/Fe] gradient is positive
($\sim 0.02$ dex pc$^{-1}$). The lower [O/Fe] in the inner region
is due to the presence
of 2G stars that are formed from gaseous ejecta with rather low
[O/Fe] from massive AGB stars.
The larger [C/Fe] in the central region is due to the higher
fraction of carbon-rich stars: stars with [C/Fe] dominates
in he [C/Fe] distribution for [Fe/H]$>-1.6$
in Figure 3.
The slope of the radial gradient of [Na/Fe] is weakly positive with
a large 1$\sigma$ dispersion ($\sim 0.3$ dex) in each radial bin.
The radial gradients of [Mg/Fe] and [Al/Fe] are rather flat with
small dispersions.
Since JP10 did not show the radial gradient of [Mg/Fe], 
it is currently not possible for the present study to discuss the consistency
of the results with observations.

There are very weak negative radial gradients for [La/Fe]
and [Ba/Fe] ($\sim -0.02$ dex pc$^{-1}$) with 1$\sigma$ dispersions
larger in the inner regions.
The radial profile of [Eu/Fe] shows a weakly positive slope, which 
is caused by a larger fraction
of 2G stars with low [Eu/Fe] that are formed
from ICM  polluted heavily by SNe II.
The radial profiles  of [Fe/H], [O/Fe], [Na/Fe], [Al/Fe],
[La/Fe], and [Eu/Fe] for
$\omega$ Cen are shown in Figure 9 of
JP10,
though the amplitudes of the slopes of the radial gradients are not described.

\subsubsection{Relations between $Y$, ${\rm [X/Fe]}$, and [Fe/H]}

As shown in Figure 6,
$Y$ and ${\rm [X/Fe] }$ depend  strongly of [Fe/H]
only for a narrow range of [Fe/H] ($-1.7 < {\rm [Fe/H]} < -1.5$).
$Y$, [C/Fe], [N/Fe], [Na/Fe], [Mg/Fe], [Al/Fe],
[Ba/Fe], and [La/Fe] increase with increasing
[Fe/H]  whereas
[O/Fe] and  [Eu/Fe] decrease with increasing [Fe/H].
The derived rapid increase/decrease of these abundances are
due largely to 2G formation from AGB ejecta mixed with SNe II ejecta,
as explained for other figures.
The simulated $Y-$[Fe/H] relation and ${\rm [X/Fe] }$ relations 
for C, N, O, Na, Mg, Al are almost  flat for 
[Fe/H] $>-1.5$: [C/Fe] and [N/Fe] very slightly decrease with
increasing [Fe/H].
Since [C/Fe] of SNe II ejecta ranges from $-0.18$ to $-0.4$ in the adopted
yields (K06),
new stars formed from ICM mixed with  SNe II  ejecta
can show lower [C/Fe].
Therefore, the decrease of [C/Fe] at [Fe/H] $>-1.5$ is due to the enrichment
of ICM
by SNe II.
 [Ba/Fe] and [La/Fe] clearly decrease with increasing
[Fe/H] for [Fe/H] $>-1.3$. Although the steady increase of [Ba/Fe] and [La/Fe]
with increasing [Fe/H] for  [Fe/H]$<-1.5$ in the simulated $\omega$ Cen
is indeed seen in observations (Figure 10 of JP10), though 
${\rm [\alpha/Fe] }-$[Fe/H] relations for the two 
are observed to be flat for [Fe/H]$>-1.5$ in JP10.
Observational studies of $\omega$ Cen by D11, however,
shows a slight  decrease of [La/Fe] with increasing [Fe/H] after
its peak around [Fe/H]$\approx -1$.

The simulated [Eu/Fe]$-$[Fe/H] relation shows a steady decrease
with increasing [Fe/H], which is not so consistent with the observed
relation with an almost constant [Eu/Fe]  over a wide range 
of [Fe/H] (JP10). 
It should be stressed here that the evolution of [Eu/Fe] with [Fe/H]
is not so clear for [Fe/H]$>-1.0$ owing to a very small number of data points
beyond [Fe/H]$>-1.0$ in JP1: it should be noted here that D11 shows
a decreasing [Eu/Fe] with increasing [Fe/H]. 
The main physical reason for the steady decrease of
[Eu/Fe] with increasing [Fe/H]
is as follows.
Although NSM ejecta can be mixed with high-density ICM polluted by AGB
ejecta,  a large fraction of the ejecta can be expelled from the 
ICM in this model so that [Eu/H] can only slowly increase.
Accordingly, SNe II ejecta, which dramatically increases [Fe/H] of ICM
if it is trapped by ICM,
can decrease [Eu/Fe] of ICM rather quickly after its mixing wit ICM.
Thus new stars formed  from ICM that is chemically polluted by SNe II to a much
lager extent than by NSM can have rather low [Eu/Fe].
The amount of AGB ejecta that can trap NSM ejecta 
($<10^6 M_{\odot}$) is so small in this forming
$\omega$ Cen that chemical enrichment by SNe II can proceed more rapidly
than that by NSMs. 
A way to reproduce the observed [Eu/Fe]$-$[Fe/H] relation better
is discussed in the next subsection on the parameter dependence.

\subsubsection{Relations between ages and chemical abundances}

Figure 7 describes the relations between ages and chemical abundances
(e.g., $Y$ and [Fe/H])  estimated over $\sim 300$ Myr 
for 11 elements in the simulated 
$\omega$ Cen. Here the ``age'' of a star in a simulation
is the difference between the formation epoch of 
the star and the final time of the simulation:  ``zero age'' means 
the youngest star in the simulation and the oldest star is 300 Myr old.
The key results of Figure 7 are as follows.
First, the age metallicity relation (AMR: age-[Fe/H] relation) 
is steep only for stars with ages older than 200 Myr,
and the AMR is quite flat for those younger than 200 Myr. 
Although stars with higher metallicities ([Fe/H]$\sim -0.5$) can be formed 
from ICM heavily polluted by SNe II of 2G stars in
the later stage of $\omega$ Cen formation in this model,
new stars with lower [Fe/H]
can be also formed from AGB ejecta not so heavily polluted by SNe II.
As a result of this,  the mean [Fe/H] for younger stars cannot be so high.

Second, steep relations between ages and chemical abundances can be 
seen for $Y$, [C/Fe], [N/Fe], and [O/Fe] for ages older than 240 Myr,
which is due to the 2G formation from AGB ejecta.
These age-abundance relations are rather flat for stars younger
than 240 Myr.
Third, age-abundance relations are  fairly flat for [Na/Fe],
[Mg/Fe], and [Al/Fe], though
[Na/Fe] depends weakly on ages for stars older than 160 Myr.
Fourth, there is not a clear trend in the age-abundance relations
in [Ba/Fe] and [La/Fe] for stars younger than 240 Myr, however,  
lower these abundances for younger ages  can be seen for stars with ages older than
240 Myr. It should be stressed here that [La/Eu] and [Ba/Eu]
for ages younger than 100 Myr are  quite high ($>0.5$, i.e.,
dominated by $s$-process elements), which is consistent
with observations by JP10,  though they are negative
for the oldest age bin.
There is a flat age-[Eu/Fe] relation for stars younger than 240 Myr,
and [Eu/Fe] is the highest in the oldest age bin corresponding to 1G stars.

\subsubsection{Chromosome maps}
Figure 8 describes the distributions of 1G and 2G stars on the [O/Na]$-$[Fe/H]
and [O/Fe]$-$[Fe/H]
maps, which can be compared with corresponding observations
(e.g., Figure 18 in JP10). These maps are similar to
``Chromosome map''
shown in  Figure 23 by  Marino et al. 2019, which describes
the observed distribution of stars on the [Na/Fe]-[O/Fe] map.
As shown in Figure 8, 1G and 2G stars are widely
distributed  on the [O/Fe]$-$[Fe/H] map and there is a bimodal distribution
for [Fe/H]$<-1.3$, which can be seen also in the observed map, though
there is a large dispersion in the observed bimodal distribution. The simulated
distribution  shows a significant fraction of stars with [O/Fe]$>0.5$ at [Fe/H]$>-1.3$,
which is not  observed. 
These stars with  large [O/Fe] (up to $\sim 1.1$) can be formed from
ICM that are polluted heavily by
massive SNe II ($m>30 M_{\odot}$).

Since K06 assumes the broadly constant Fe yield, irrespective of the progenitor mass,
[O/Fe] in the ejecta of massive SNe II is quite large.
If other O and Fe yields from other papers (e.g., [O/Fe]=0.5 from Tominaga et al. 2007)
are adopted,
the fraction of such O-rich star can be reduced to the observed level.
In order to demonstrate this point,
a comparative model M1a with [O/Fe]=0.5 in massive SNe II is investigated,
and the results are shown in Figure 8.
Clearly, the level of consistency between the observed and simulated 
distributions of stars on this map is significantly higher in M1A than
in M1.
This result implies that
it might be better for our future works on the chemical
evolution of $\omega$ Cen  to
consider the Fe yields implied by the light curve analysis of supernovae,
which indicates an increasing Fe yield with increasing  progenitor’s mass
(e.g., Tominaga et al. 2007).

1G and 2G stars are distributed on two narrow ``stripes'' on the [O/Fe]$-$[Fe/H]
map for [Fe/H] $<-1.3$ whereas the observed bimodal distribution does not show
a clear gap between the two groups ([O/Fe]$\approx -0.4$ and $\approx 0.4$).
Since the natal GMC of $\omega$ Cen is assumed to have no [Fe/H] and [O/Fe] spreads,
the 1G stars formed from the gas cannot show a wider distribution on the map.
If such initial abundance spreads within a GMC is considered in future simulations,
the observed distribution will be able to be better reproduced.
Although the simulated distribution of stars on the [O/Na]$-$[Fe/H] map
shows a bimodality for [Fe/H]$<-1.3$, the overall
distribution  appears to be dissimilar to the observed one owing to the presence
of stars with high [O/Na] ($>1$) at [Fe/H]$>-1.3$.  
The observed distribution appears to show two major  groups of stars, one with
[O/Na]$\approx -1$ and the other with [O/Na]$\approx 0.4$, though
there is no clear gap between the two groups.
The group with lower [O/Na] in the simulated distribution has a bit too high
[O/Na], which is again due to the adopted large [O/Fe] from massive SNe II.
Intriguingly,
the observed distribution has a very minor fraction (2 out of 29 stars;
6.9\%) of stars with [O/Na]$>0$
and [Fe/H]$>-1$ (``zone of avoidance''). This almost empty area on this map
cannot be seen in the simulated distribution, which is an inconsistency
between observations and the present simulation.

The origin of this apparent zone of avoidance is yet to be understood. 
First of all, it is not clear why this can be seen for a higher metallicity
range ([Fe/H]$>-1$) only. Clearly a large fraction of stars at [Fe/H]$<-1$ shows
high [O/Na] ($>0.5$) in the observed $\omega$ Cen, which is consistent with
new star formation from gas polluted by SNe II ejecta. The lack of such
stars at [Fe/H]$>-1$ could be therefore due to a much smaller degree of
chemical pollution of ICM by SNe II ejecta with high [O/Na]
(high [O/Fe])  for that metallicity range.
One possibility is that ICM is polluted only by low-mass SNe
for [Fe/H]$>-1$. It is beyond the scope of this paper, however,
to discuss why [Fe/H]$\approx -1$ is the ``turning point'' when the chemical 
enrichment processes by SNe II can significantly change: this should be one
of the key issues in our future studies.
It should be noted here  that the fraction of stars with [O/Na]$>$ and [Fe/H]$>-1$
is only 7.8\%, which is only slightly (by 13\%) higher than the observed
value (6.9 \%). Therefore, it could be possible that if 
[O/Na] and [Fe/H] are investigated for a much larger number
of stars ($\approx 10^4$), then the above zone of avoidance disappears.

Figure 9 describes the distributions of 1G and 2G stars on the [N/Fe]$-Y$ map
in the simulated $\omega$ Cen, which shows (i) one strong peak due to
1G stars and (ii) two elongated and wide distributions along [N/Fe]-axis
for $Y\sim 0.31$ and $Y\sim 0.35$. 
Clearly,  He-rich stars can show a wide range of [N/Fe] and
are more likely to have higher [N/Fe]. The derived wide range of [N/Fe] for 
a given $Y$ is due to 2G formation from AGB ejecta with higher $Y$
and higher [N/Fe] that is mixed
with SNe II ejecta with lower [N/Fe] ($<-0.8$).
Bellini et al. (2017) have recently revealed that although
``bMS'' stars (i.e., those on
the blue main sequence on the CMD) have rather high $Y$ and [N/Fe],
some stars (``MSd'' population)
have moderately high $Y$ and moderately high [N/Fe] (see their Table 1).
These observed populations with different $Y$ and  [N/Fe] are qualitatively
consistent with the results in Figure 9.
They have also identified 15 distinct subpopulations based on [Fe/H], $Y$, and
[N/Fe] (see their Table 1), which are not clearly seen in Figure 9.
The lack of multiple distinct peaks in  Figure 9 reflect the fact that star formation
is more continuous in the formation of 2G stars in the present study.

\subsubsection{[La/Eu]-[Fe/H] relation}
One of the most intriguing observational results of $\omega$ Cen is that
it shows a large [La/Eu] ($>0.1$) for  [Fe/H] $>-1.2$ (e.g., S00, JP10, D11).
Such a large [La/Eu] cannot be seen in halo field stars of the Galaxy (JP10),
which implies that the chemical evolution of $\omega$ Cen is quit different
from that of the Galactic halo.
Since this [Eu/La] evolution depends strongly on the ejection rates of gas
from AGB stars, NSMs,
and SNe II,  the observed trend with [Fe/H] can be a strong constraint on
the formation of $\omega$ Cen (JP10). As shown in Figure 10,
[Eu/Fe] rapidly increases with increasing [Fe/H] and reaches its peak ($\sim 0.6$)
around [Fe/H]=$-1.5$. 
This rapid increase reflects the fact that AGB ejecta can be much more
efficiently retained in  ICM than NSM ejecta in $\omega$ Cen.
Accordingly,
these different degrees of retention capabilities of AGB and NSM ejecta
in $\omega$ Cen formation can distinguish it from other GCs and 
halo field stars.
The derived  rapid increase is consistent with the observed trend
by JP10, however, the location of the peak ([Fe/H]$\sim -1.5$) is a bit too
metal-poor in comparison with the observed location ([Fe/H]$\sim -1$; D11).
This apparent inconsistency implies that the chemical enrichment of the ICM
by AGB stars
in the simulated $\omega$ Cen can proceed a bit too rapidly.

The simulated very slow decline of [La/Eu] with increasing [Fe/H]
can be seen in the observed [La/Eu]$-$[Fe/H] relation (e.g., Figure 18 in JP10).
Also, it should be stressed that observations show a minor fraction of 
stars with [La/Eu]$<0$ for [Fe/H]$>-1.2$, which is also seen in the simulated
$\omega$ Cen. The low and positive [La/Eu] ($\sim 0.2$) 
 at [Fe/H]$\sim -0.5$ in the simulated
$\omega$ Cen is consistent with observations by JP10 and D11,
which strongly suggests that this fiducial model can properly describe
some key processes (but not all)  in the 
mixing  of AGB, NSM, and SNe II ejecta. Although
observations do not clearly describe the dispersion of [La/Eu] for a given
[Fe/H],  the derived dispersions in this model appears to be 
compatible with the observed ones (JP10).
It should be stressed that the adopted La yield is  by a factor of 2.5 larger
than that shown in F14. As discussed later,  the models with lower La yields
show  low [La/Eu] ($\le 0$) at [Fe/H]$\sim -0.5$
that is not consistent with observations.

\subsection{Parameter dependence}

It is confirmed that  some properties of the simulated
$\omega$ Cen depend strongly on the input parameters such as
the degree of initial rotation of gas and the mixing process of 
AGB and NSM ejecta, though others do not.
We here describes the key physical properties of $\omega$ Cen that depends
strongly on model parameters in Figures 12$-$15.

\subsubsection{Structure and kinematics of He-poor and He-rich populations}

All of the present models with different parameters show
a strong central concentration of He-rich stars in the simulated
$\omega$ Cen:  the 3D structure of He-rich stars do not depend
on model parameters,
$M_{\rm gmc}$,  $R_{\rm gmc}$, and mixing processes of AGB and NSM ejecta.
However, the projected 2D kinematics of He-poor and He-rich stars
in the simulated $\omega$ Cen  depend strongly on $f_{\rm rot}$
(i.e., the initial fraction of rotational energy in kinetic energy 
for the natal GMC).
Figure 11 shows that the amplitudes of rotation for He-poor and 
He-rich populations are significantly enhanced in the model
M2 with $f_{\rm rot}=0.03$. 
The maximum amplitude of rotation ($V_{\rm max} \approx 5$ km s$^{-1}$) in this
M2 is more consistent with observational studies of line-of-sight 
velocity profile of $\omega$ Cen by Bianchini et al. (2013),
which showed $V_{\rm max} \approx 6$ km s$^{-1}$ around $R=500$ arcsec.
It is confirmed that the amplitudes
of rations of the two populations are higher in the models
with larger $f_{\rm rot}$. 
However, the rotational amplitude is higher in the He-poor population
than in the He-rich population for the simulated $\omega$ Cen in these
models. The models with larger $f_{\rm rot}$ show lower conversion efficiencies of
gas to new (2G) stars ($<0.5$) and the total mass of 2G stars in the models
with $f_{\rm rot} \ge0.05$ are too low ($<10^5 {\rm M}_{\odot}$) to be consistent
with the observed mass of He-rich stars (Norris 2004): this point is discussed
in 3.2.4.

\subsubsection{Radial abundance gradients}

As shown in Figure 12,
the slopes of radial abundance gradients (negative or positive)
do not depend on model
parameters for all but [Eu/Fe], though the details of the profiles
(the central value) are different between different models.
This is mainly because the formation of 2G star from AGB ejecta
mixed with SNe II and NSM ejecta does not depend strongly on
the model parameters.
The positive slopes of [Eu/Fe] derived in some models
are caused by the too large [Eu/Fe] of 2G stars formed from ICM 
polluted efficiently by NSM in the inner regions of these models.
Since the details of mixing processes of NSM ejecta are not understood
for ICM, the present study cannot make a robust conclusion as to whether
the [Eu/Fe] radial gradient of simulated $\omega$ Cen should be negative or 
positive. It should be noted here that the observed [Eu/Fe] gradient
appears to be  weakly negative (Figure 9 of JP10),
within the central 10 arcmin.
 
\subsubsection{[X/Fe]$-$[Fe/H] relations}

Although ${\rm [X/Fe]}-$[Fe/H] relations are not so different between
different models for X=C, N, O, Na, Mg, and Al,
they depend strongly on the mixing process of NSM ejecta for
Ba, La, and Eu, simply because they are produced efficiently by NSM.
Figure 13 describes how  the simulated [Eu/Fe]$-$[Fe/H]  relation
depends on $r_{\rm nsm}$ (search radius of NSM),
$r_{\rm s,0}$ (stopping length), and $f_{\rm ret}$ (retention fraction
of NSM ejecta). As shown  in Figure 13 for
the models with uniform mixing of NSM ejecta,
 if all NSM ejecta is retained ($f_{\rm ret}=1$) in a GC,
then [Eu/Fe] can become too large ($>2$) to be consistent with
the observed [Eu/Fe].
Clearly, the simulated [Eu/Fe]$-$[Fe/H] relation is more consistent
with the models with small $f_{\rm ret}$
(i.e., partial retention)  than with those with $f_{\rm ret}=1$.
It should be noted here, however, that all NSM ejecta
is assumed to be mixed well with  ISM
with a mass of $\sim 10^6 M_{\odot}$ in theoretical
studies for [Eu/Fe] evolution in galaxies (e.g., TS14).
Such a 100\% retention rate of NSM ejecta within ICM
is highly unrealistic in forming GCs owing to the small amount of ICM
($<10^6 M_{\odot}$).
Thus,  it is not a problem
that the models with 
small $f_{\rm ret}$ can better reproduce the observation.
In the density-dependent (``D'')
 models with larger $r_{\rm nsm}$ and larger $l_{\rm s,0}$,
NSM ejecta can be mixed more uniformly for a wider region
of ICM
so that the net retention rates  of NSM ejects can become higher.
Accordingly, these models do not show a strongly negative [Eu/Fe]
at higher [Fe/H] ($>-0.8$) that is inconsistent with
the observed slightly positive  (JP10)
or slightly negative  [Eu/Fe] (S00, O11) for [Fe/H]$>-0.8$.

As shown in Figure 14,
the models with larger yields of La can better reproduce
both the observed [La/Fe]$-$[Fe/H] and [Eu/Fe]-[Fe/H] relations
for [Fe/H]$>-0.8$.
Since there is an uncertainty in theoretical predictions of
chemical yields for La, 
it would not be unreasonable for the present study to
use larger La yield for reproducing better the observations.
Figure 14 also demonstrates that such models with larger
yields of La 
can better explain the observed
positive [La/Fe] at higher [Fe/H] ($ \approx -0.5$).
It should be noted here that the models with Y1 yield (i.e., lower
La yield) 
do not show positive [La/Fe] at [Fe/H]$ \approx -0.5$ owing
to the small [La/Fe] at [Fe/H]$>-0.5$. 
The model with Y2 yield also show a  small ($\approx 0$) [La/Fe] at 
[Fe/H]$\sim -0.5$, which is not so consistent with observations (JP10).
These low [La/Eu] at higher [Fe/H] can be seen for [Ba/Eu] in these models.
These results strongly suggest that Ba and La yields
should be larger than the predictions by F14 to reproduce the observed
[Ba/Eu] and [La/Fe] at higher [Fe/H] in $\omega$ Cen.

\subsubsection{Star formation efficiency for He-rich stars}

The observed total mass of He-rich stars ($10-20$\% of its total stellar mass
; Norris 2004) can be used as a key constraint on the IMF and the original
mass of $\omega$ Cen (Bekki \& Norris 2006).
Although the fiducial model M1 shows that the total mass of He-rich stars
with $Y>0.25$ ($M_{\rm 2g}$)  is $\sim 8 \times 10^5 M_{\odot}$, which is consistent
with observations,  some other  models with different $M_{\rm gmc}$,
$R_{\rm gmc}$,  and $r_{\rm gmc}$
do not show
such a large mass owing to the lower conversion (star formation)
efficiency of the ICM (M13$-$M22).
 The models with larger $f_{\rm rot}$ ($ \ge 0.05$) show
$M_{\rm 2g} < 10^5 M_{\odot}$, because the the ICM has  more extended
spatial distributions and thus  lower surface gas densities
these models with larger initial angular momentum.
This result combined with those of stellar kinematics depending on $f_{\rm rot}$
can allow us to narrow down the range of $f_{\rm rot}$ that can explain
both the observed $M_{\rm 2g}$ and $V_{\rm max}$.

Furthermore, $M_{\rm 2g}$ can be lower than $10^5 M_{\odot}$
in the models with lower $M_{\rm gmc}$
($<10^7 M_{\odot}$ for $f_{\rm rot}=0.01$). 
This result implies that there is a threshold GMC mass 
($M_{\rm gmc} \ge 10^7 {\rm M}_{\odot}$) above which $\omega$
Cen can be formed with the right mass of 2G stars.
The low-density GMC  model with larger $R_{\rm gmc}$ ($=280$pc) also shows
a lower $M_{\rm 2g}$ ($=5.0 \times 10^5 {\rm M}_{\odot}$ for $f_{\rm rot}=0.01$),
which is lower than the observed mass of $Y$-rich stars.
The model with $r_{\rm gmc}=500$ pc (i.e., located in the outer part
of the dwarf's galaxy) shows $M_{\rm 2g}=3.3 \times 10^5 M_{\odot}$
that is significantly lower than that in the fiducial model with $r_{\rm gmc}=0$.
It is confirmed that GMCs with such larger $r_{\rm gmc}$ show
low star formation efficiencies in 2G formation.
This suggests that the star formation efficiency from ICM in a GC can be higher if 
the GC is located in the nuclear region of its host dwarf.
The main reason for the higher star formation efficiency for $r_{\rm gmc}=0$
is that AGB ejecta can be more efficiently retained in the GC if it is located
in the nuclear region.

\section{Discussion}

\subsection{Short vs long formation timescales}

The formation timescale of $\omega$ Cen in the present scenario ($\sim 300$ Myr) is
much shorter than those (more than a few Gyr)  suggested by previous observational studies,
which used 
different photometric and spectroscopic methods to derive
the ages range: see Table 1 of the present study and 
Table 5 of Stanford et al. (2006) for
a list of age ranges estimated for stellar populations
of $\omega$ Cen 
in previous work done before 2006.
The new short formation timescale can explain
a number of the observed chemical abundance patters of $\omega$ Cen, 
as demonstrated in \S 3.
Furthermore, it is consistent with the observed low [Mn/Fe] at
[Fe/H]$\sim -1.0$ observed in $\omega$ Cen
(Cunha et al. 2010).
However, it has not been clearly explained in the present study how
star formation can be almost truncated $\sim 300$ Myr after its formation.
Massive $\omega$ Cen can have numerous AGB stars with lower masses
($\le 3 M_{\odot}$)
even when it becomes older than  300 Myr.
These AGB ejecta can be easily trapped by 
$\omega$ Cen (and by its host dwarf), and consequently
it can be converted into new stars.
However, if such AGB ejecta with rather high [Ba/Fe] and  [La/Fe]  
(e.g., [La/Fe]$\sim 1.6$ and $2.1$ for $m=3 M_{\odot}$ and  $2M_{\odot}$,
respectively;  F14 )
is converted into new stars without dilution by pristine
gas, then
[Ba/Fe] and [La/Fe] can be too high to be consistent with
observations.

Therefore, there should be a mechanism by which star formation
can be truncated $\sim 300$ Myr after the formation of $\omega$ Cen.
There are possible explanations for the origin of the  sudden truncation of 
star formation.
One is that SNe Ia in the central region of $\omega$ Cen can expel
all of the remaining ICM to truncate further star formation:
this idea was already discussed in D'Ercole et al. (2008).
Since a fraction of SNe II ejecta can be retained by the central region
of $\omega$ Cen's host dwarf in the present models,
this idea has need to explain why SNe Ia, whose explosion energy
is similar to that of SNe II, can expel all of the remaining ICM
within $\omega$ Cen.
It seems that this idea is not so promising owing to the lack of
a possible convincing mechanism for the  removal of gas
preferentially by SNe Ia (not by SNe II).

An alternative idea is 
that the host dwarf galaxy of $\omega$ Cen experiences strong
ram pressure stripping of its ICM by the warm/hot halo gas
of the Galaxy $\sim 300$ Myr after the formation
of $\omega$ Cen.
It is not clear, however, why ram pressure stripping can 
suddenly become so efficient to remove the ICM around 300 Myr after
the formation of $\omega$ Cen.
One possible explanation is that  $\omega$ Cen
is formed in the first pericenter passage of its host dwarf with respect
to the Galaxy (e.g., owing to rapid gas infall during galaxy interaction
between the two)  and then it loses ICM through ram pressure stripping
in the second pericenter passage that is   $\sim 300$ Myr after
the first one.
In this explanation,  the pericenter distance of the host dwarf 
in the later multiple pericenter passages (2nd, 3rd, etc) should be
significantly smaller than that of the first one so that ram pressure
stripping is only effective in the later pericenter passages.
If the host dwarf is massive enough to experience dynamical friction
of the Galactic dark matter halo, then such a decreasing pericenter distance
with increasing number of pericenter passages  could be possible.

The main reason for the proposal of the short formation timescale
of $\omega$ Cen is that there is no clear decrease of ${\rm [\alpha/Fe] }$
with [Fe/H] in observations (e.g., JP10). 
Although the number of stars used in JP10 (855) is large enough,
a larger number of stars with [Fe/H]$>-0.8$ is required to confirm
that there is no decrease of [Ca/Fe], [Si/Fe], 
and [Ti/Fe] with increasing  [Fe/H]  
for [Fe/H]$>-0.8$,
because the number of stars for that metallicity range
is rather small in JP10.
The decreasing trends of [Ba/Fe], [La/Fe], and [Eu/Fe] with increasing
[Fe/H], which are the prediction of the present scenario,
need to be confirmed in future observations with a much larger
number of the investigated stars.
It is beyond the scope of this paper how we can reinterpret the observational
results of color magnitude
diagrams (CMs) of $\omega$ Cen
suggesting the long formation timescale ($>2$ Gyr) in the context
of the present scenario. A combination of photometric errors and differences
in [Fe/H], $Y$, and CNO abundances among stars could possibly broaden the distributions
of the stars on the CM diagram thus mimic the large age spread.

\subsection{Dependence on AGB yields}

The present study has investigated the chemical abundances of
stars in $\omega$ Cen based on only one AGB model (F14). It is therefore
possible that the present results can be  changed if
AGB yields from other groups are incorporated into the adopted
chemodynamical simulations of GC formation.
We here discuss this point using the AGB yields from Ventura et al. (2013, V13),
which have been used in chemical evolution studies of GCs with
multiple stellar populations. 
One of  differences between F14 and V13
is [O/Fe] in stellar winds of 
massive AGB stars ($m \ge 6 {\rm M}_{\odot}$):
it is $\approx -0.6$ and $-0.8$ 
in F14 and V13, respectively. 
V13 also showed that the level of Mg-depletion in massive AGB stars
with $m \approx [5-6] M_{\odot}$ ([Mg/Fe]$\approx -0.2$)
is  higher than that predicted in F14 ([Mg/Fe]$\approx 0.4$).
Furthermore, [N/Fe] in stellar winds of
intermediate and massive AGB stars ($m \ge 5M_{\odot}$)
are lower in V13 (ranging from 1.4 to 1.6) than in F14
(from 1.8 to 2.5). Although C-depletion in ejecta from  AGB
stars with $m \ge 6 M_{\odot}$  can be clearly seen in V13
($-0.51 \le$[C/Fe]$\le -0.27$),
it is not seen in F14: [C/Fe] is as high as $\approx$ 1 for such a mass
range.

These differences suggest that the following properties of
the simulated $\omega$ Cen can change,
if the AGB yield table by V13 is adopted in our chemodynamical simulations
of GC formation.
First,  the mean [O/Fe] of 2G stars in the simulated
$\omega$ Cen can be  lower than that derived in the present study,
because the ICM is polluted by winds from massive AGB stars with
smaller [O/Fe].
This  means that the simulated [O/Fe]$-$[Fe/H]
and [O/Na]$-$[Fe/H] relations (shown in Fig. 8) can be 
better fit to the corresponding observations (JP10).
It is, however, clear that the observed stars with [O/Fe]$\approx -1$ (JP10)
cannot be simply explained, even if V13 is adopted.
Second, the radial gradient of [O/Fe] can be slightly steeper 
owing to the lower [O/Fe] in the central 2G population of the simulated $\omega$ 
Cen.

Third, the location of the peak in the [Mg/Fe] distribution
can be shifted toward a lower [Mg/Fe]
(by $\approx 0.2$ dex at most)  owing to the lower mean [Mg/Fe] of 
2G population. 
Fourth, the number fraction of stars with high [N/Fe] ($>1$) can be 
reduced owing to the lower [N/Fe] of AGB ejecta in V13.
Norris \& Da Costa (1995) investigated
the chemical abundances of  40 red giants
in $\omega$ Cen and found [N/Fe]$<1$ for almost all of them.
Although Marino et al. (2011) found that
stars with [Fe/H]$>-1.5$ have  
[N/Fe]$\approx 1.5$, there are no stars with [N/Fe]$>1.8$.
It is therefore possible that
our future simulations with AGB yields from V13  ([N/Fe]$<1.6$) will be able to 
reproduce better the observed [N/Fe] abundances.

Fifth, the number fraction of stars with [C/Fe]$>0.5$ among 2G stars can be 
decreased.   
The simulated $\omega$ Cen has a non-negligible fraction
of stars with [C/Fe]$>0.5$, which is not so consistent with
the observational results by Norris \& Da Costa (1995)
and Marino et al. (2011), which showed
almost all of the investigated stars show [C/Fe]$<0.5$. 
Accordingly,  adoption of AGB yields by V13 can improve 
the level of consistency between the observed and simulated [C/Fe]
distributions.
Sixth, $Y$ distributions and
radial $Y$ gradients cannot change significantly,
because the predicted $Y$ in AGB winds is not so different between F14 and V13. 
The lack of AGB yield table for
[La/Fe] and [Ba/Fe] 
in V13 does not allow us to discuss how these abundance patters can be
changed in our chemodynamical simulations with the yields from V13.

\subsection{The origin of global rotation}

The present  study has shown,  for the first time,
that (i) both He-poor and He-rich stars in $\omega$ Cen can have
global rotation with respect to its center
and (ii) the rotation amplitude can be higher in the He-poor population.
Our previous studies of GC formation, which do not include
SN feedback effects of 2G stars on ICM, showed the higher amplitude
of rotation in 2G stars that correspond to He-rich populations.
Therefore, the above result (ii) implies that
the number fraction of SNe II among 2G stars can possibly
determine the stellar kinematics of 2G populations in GCs.
Since $\omega$ Cen is massive enough to form high-mass stars
that can explode as SNe II,
the stellar kinematics of 2G stars can be significantly influenced
by SNe II feedback effects (i.e., stirring of their host ICM)
to have the lower amplitude of global rotation.
On  the other hand, GCs with lower masses cannot form SNe II progenitor
stars so that the 2G stars can have more rotation than 1G stars
owing to their dissipative formation from ICM.

The present study has also shown that the rotational energy of the
natal GMC of $\omega$ Cen can control the present-day rotation
amplitude of He-poor (1G) stars in $\omega$ Cen.
The rotational kinematics of stars in typical GCs
can be significantly changed during
the long-term dynamical evolution of GCs driven by
two-body relaxation  (e.g., Mastrobuono-Battisti \& Peters 2016).
Although the long-term dynamical evolution of such a massive GC with
at least two distinct populations with different kinematics
is yet to be investigated (e.g., for GCs with lower masses),
it cannot be so much changed for $\omega$ Cen with a long
relaxation timescale.
Therefore, it is reasonable to conclude that the observed
$V/\sigma$ in $\omega$ Cen can be used to infer the original
$f_{\rm rot}$ of its natal GMC.
Thus it is our future study to investigate whether
the formation of a GMC with the suggested $f_{\rm rot}$ is 
possible in a dwarf galaxy at high $z$.

\subsection{Was $\omega$ Cen really formed in a dwarf galaxy's center ?}

It has long been considered that $\omega$ Cen was the nucleus of a dwarf
galaxy that had been completely destroyed by the Galaxy
(e.g., Freeman 1993). The observed large metallicity spread
of $\omega$ Cen was also considered to be observational evidence that supports
this scenario, and the metallicity spread was 
suggested to be caused by repetitive gas infall to the nucleus
over several Gyr in the early evolution of the host dwarf (e.g., BF03).
As has been demonstrated in the present study,
the origin of the observed wide range of [Fe/H] could not be due to
the discrete star formation events due to repetitive gas infall to the nuclear
region of $\omega$ Cen's host dwarf 
over several Gyr suggested
by BF03.

Formation of $\omega$ Cen in the nuclear region of its host dwarf
has the following advantages in explaining the observed
chemical abundances of $\omega$ Cen.
First, the star formation efficiency from AGB ejecta can be
quite high ($\sim 0.5$), because almost all  gas in AGB winds can be retained
and consequently used for secondary star formation owing to the deep
gravitational potential well.  Second, a significant amount
of  ICM polluted by SNe II can be
also retained in the nuclear region so that it can be converted
into new stars with higher [Fe/H]. Third, the stellar system composed largely
of 1G stars does not lose the stars that can become AGB star to eject
gas, because it does not suffer tidal stripping by its host.
This can end up with the larger mass of 2G stars from AGB ejecta.
Fourth, the existing field stars of the host galaxy
could be gravitationally trapped by $\omega$ Cen to avoid tidal stripping
during the final destruction of its host dwarfs.
The field stars with low metallicities can be observed as the
stars with [Fe/H]$<-1.8$ within $\omega$ Cen.

Thus, several properties of chemical abundances characteristic
for $\omega$ Cen can be closely related to its formation in
the nuclear region of a dwarf galaxy orbiting the early Galaxy.
Such unique formation processes of $\omega$ Cen
could also explain other physical properties of $\omega$ Cen.
For example,
the largest GC among the Galactic GC can be also explained by
its formation in the host's nuclear region, where a large amount of
gas can be transferred to form a very massive GMC.
A possible disadvantage of $\omega$ Cen formation in the nuclear
region of its host is that star formation can continue in the
deep potential well more than 300 Myr.
A  number of recent observational studies
using the latest Gaia DR2 data have successfully
discovered the tidal streams that can be associated with $\omega$ Cen
(e.g., Ibata et al. 2019).
It is our future study to construct a model which can reproduce
not only the chemical and dynamical properties of $\omega$ Cen
but also these latest proper motion results of the host dwarf galaxy.

\subsection{Can the IMF of 2G stars
in $\omega$ Cen distinguish it from other GCs ?}

The present study has revealed that there are two unique physical processes
that distinguish $\omega$ Cen from other ``ordinary''
GCs with no/little [Fe/H] spread.
One is the mixing of  SNe II ejecta  of 2G stars 
with AGB and  NSM ejecta of 1G stars in the central region of $\omega$ Cen.
The other is the  efficient
retention of such mixed ejecta in the central region for
prolonged star formation. 
In order to form SNe II in 2G formation,
the upper mass cut-off ($m_{\rm upp}$) should be 
larger  than $8 M_{\odot}$ in the 2G formation of $\omega$ Cen.
Our previous one-zone
models of GC formation showed that 2G star formation can continue
for $\sim 2 \times 10^8$ yr,  if the IMF of stars 
in 2G formation is top-light with
the upper mass cut-off ($m_{\rm upp}$) less than $8 M_{\odot}$
(Bekki et al. 2017, BJK17). This is mainly because
2G formation cannot be truncated by the energetic feedback effects of SNe II
(i.e., no SNe II).
The lack of chemical enrichment by SNe II in 2G formation ends up with
no [Fe/H] spreads among 2G stars.

Accordingly, it is possible that the IMF in 2G formation
can be a crucial factor that distinguishes between $\omega$ Cen
and other ordinary GCs with multiple stellar populations yet no/little
[Fe/H] spreads. In such a massive GC as $\omega$ Cen, 
$m_{\rm upp}$ can be larger
than $8 M_{\odot}$ owing to the higher star formation rate
of 2G stars (BJK17): however, it should not 
be so large ($<20 M_{\odot}$)
so that the mass fraction of SNe II ($f_{\rm sn2}$) can be low to have 2G stars
with the maximum [Fe/H] as low as  $-0.3$: high $f_{\rm sn2}$ can end up
with the formation of new stars with [Fe/H]$\approx 0$, which is not 
observed.
On the other hand, 
$m_{\rm upp}$ can be 
lower than $8 M_{\odot}$ owing to the lower star formation rate
of 2G stars in ordinary GCs.
Therefore, the mass-density of ICM, which can determine the IMF of 2G stars
(e.g., Marks et al. 2012),
can be a key factor that distinguishes between $\omega$ Cen and 
other ordinary GCs.
The ICM of forming $\omega$ Cen can possibly have a high mass-density
owing to (i) its original large mass and (ii)
its formation in the central region of its host dwarf.

\subsection{Unresolved problems and future works}

Although the present study has demonstrated that some of the observed
properties of $\omega$ Cen can be explained by the new model with
the short formation timescale,
it cannot explain a number of key observations of $\omega$ Cen in a fully self-consistent
manner.
First,  it is not clear why there are 15 distinct populations 
with possibly different $Y$, [N/Fe] and [Fe/H] in $\omega$ Cen
(see Table 1 in Bellini et al. 2017).
The present study has shown that there are three peaks 
in the $Y$ distribution of the simulated $\omega$ Cen,
which means that the GC has three distinct populations.
Although the simulated $\omega$ Cen has a wide range 
of [Na/Fe],   several distinct peaks in the [N/Fe] have not been
found in the present study, which is not so consistent with
observational results by Bellini et al (2018).
The star formation processes in the present study are not discrete events
so that the observed large number of distinct
populations cannot be simply explained by the present models.
It has been demonstrated that 
such discrete star formation events can be
possible if each star formation event is completely truncated by
its SNe II  (Bekki et al. 2017). However, such complete truncation
of star formation does not occur in the present models.

Second, the mass-ratio of metal-poor 1G ([Fe/H]$\approx -1.7$) to  metal-rich 2G
stars ([Fe/H]$>-1.5$) is a bit too large ($>10$): the simulated [Fe/H]
distribution has a too strong peak around [Fe/H]$\sim -1.7$
whereas the observed one shows multiple distinct peaks (e.g.,
Figure 7 in JP10).
The inability of the present models
to reproduce the observed significant fraction  of more metal-rich stars 
is due partly to the large original mass of 1G stars in the simulated $\omega$ Cen
($\sim 10^7 M_{\odot}$).
Although $\omega$ Cen can lose a significant fraction of its 1G stars
during its tidal interaction with the Galaxy,
it can lose at most 70\% of its original mass to match the observed
present-day total mass ($\approx 3 \times 10^6 M_{\odot}$ in 1G).
Therefore, other mechanism(s) is required to explain the large fraction
of 1G stars that are observed to be centrally concentrated 
within 10 arcmin of $\omega$ Cen.
One possible idea is that the IMF of 1G stars from its natal GMC
is top-heavy to increase the number of AGB stars thus the total
mass of AGB ejecta from which 2G stars can be formed. Such a top-heavy IMF can also
decrease the number of low-mass stars ($m<0.8 {\rm M}_{\odot}$)
that can be observed in the 
present-day $\omega$ Cen.  Recent theoretical models for IMFs
have shown that the IMF in GC formation should be top-heavy
to explain the mass function of low-mass stars in GCs
(e.g., Marks et al. 2012).
Therefore, this idea of top-heavy IMF is promising and worth an
investigation in our future study to resolve this problem.

Third,  the details of the mixing processes of NSM ejecta is not fully
understood in the formation of $\omega$ Cen. There are still two unknown
parameters that describe the processes, i.e., the stopping length
($l_{\rm s}$) and the retention fraction of the ejecta ($f_{\rm ret}$).
Although the present study has demonstrated that some models
with a particular range of these two parameters
can reproduce the observed [Eu/Fe]$-$[Fe/H] relation,
the physical processes that control $r_{\rm s}$ and $f_{\rm ret}$
are yet to be understood fully. We will need to perform
high-resolution hydrodynamical simulations that
are specially focused on the evolution of ICM surrounding one NSM event
in order to quantify how the two parameters depend on
the physical properties of ICM such as the density, total mass,
and magnetic field.
Fourth, the details of the internal stellar kinematics of $\omega$ Cen
are yet to be fully explained by the present study, which has discussed
only the global rotation of He-poor and He-rich populations.
The velocity dispersion dependent on stellar masses
and the orbital anisotropy of stars in $\omega$ Cen have been recently
revealed (e.g., Bellini et al. 2018; Jindal et al. 2019).
Since the present study did not investigate the long-term dynamical
evolution of $\omega$ Cen, it cannot discuss these latest observations
in a quantitative way.
It is accordingly our future dynamical study to reproduce these
observations in the context of the present new scenario of $\omega$ Cen.

Finally, the observed low [Rb/Zr] ($\sim -0.5$) over a wide range of [Fe/H]
in $\omega$ Cen (e.g., S00)
cannot be simply explained by the present model with
a short formation timescale.
Such a low [Rb/Zr] of $-0.5$ in S00  was suggested to be
consistent with chemical enrichment by low-mass AGB stars ($m<3 M_{\odot}$) 
and thus with a long star formation timescale.
Indeed the latest stellar yields of [Rb/Zr] for AGB stars from Karakas et al. (2018) 
are larger than $-0.063$ for $m>4 M_{\odot}$, which is significantly larger than
the observed values by S00. 
There are two possible ways to solve this discrepancy in the context
of the present short formation timescale.
One is to consider that 
low-mass AGB models with enhanced $Y$,
which correspond to 2G stars in the present study,
 will evolve more than twice as fast
(e.g., Karakas et al. 2014).
Accordingly, it is possible that 2G He-rich low-mass AGB stars ($m\sim 2.4 M_{\odot}$)
whose lifetimes can be as short as $200$ Myr can enrich ICM with their ejecta with
low [Rb/Zr] even in the $\sim 300$ Myr formation timescale of $\omega$ Cen.  
New stars formed from such ejecta from He-rich low-mass AGB stars (mixed with
SNe II ejecta) can show low [Rb/Zr], though it is not clear whether the observed
 low [Rb/Zr] over a wide range of [Fe/H] can be reproduced.

The other is that the field low-mass AGB stars in the $\omega$ Cen's host dwarf
galaxy can chemically pollute the ICM of $\omega$ Cen owing to the trapping of
the gaseous ejecta by the dwarf itself. 
In this idea,  the $\omega$ Cen's host GMC was formed well after the formation
of its host dwarf galaxy. This idea is promising, as long as
the AGB ejecta of the field stars can be retained within the central $<30$ pc 
of $\omega$ Cen (i.e., not just by the dwarf itself). This means that if the dwarf
galaxy is a rotating disk, then the gas ejected from the field stars need to 
lose a large amount of angular momentum so that it can be funneled to the inner
$<30$ pc of $\omega$ Cen (classic ``fueling'' problem).
If the dwarf is a spheroid dynamically supported by random motion
of stars, then such a fueling problem would not occur.
Furthermore, 
the right amount of gas ejected from the AGB stars needs to be required to
lower [Rb/Zr] from 0 in the ICM to the observed $-0.5$.
The required
large amount of  ejecta of low-mass AGB stars
with normal $Y$ can lower $Y$ of ICM to  end up with the formation of 2G stars
with normal $Y$ $-$ which would not be consistent with high $Y$ of 2G stars
in $\omega$ Cen. Thus the idea of short evolution of He-rich AGB stars
appears to be more promising and thus worth a detailed investigation
in our future studies.
It should be finally noted here
that the average [Rb/Zr] for three stars of $\omega$ Cen observed by
Vanture et al. (1994)  is  0.2.

\section{Conclusions}


We have adopted a new formation scenario of the Galactic
GC  
$\omega$ Cen 
and thereby investigated the chemodynamical evolution of $\omega$ Cen
using our original hydrodynamical simulation code with
chemical enrichment and feedback effects from Type II supernovae
(SNe II) and  asymptotic giant branch (AGB) stars and a new retention model
of ejecta from  neutron star
mergers (NSMs).
In the scenario, $\omega$ Cen 
was formed from a giant molecular
cloud (GMC) in the central region of an ancient dwarf galaxy that
was accreted onto the Galaxy. 
The nuclear GC could become $\omega$ Cen after its host galaxy was completely
destroyed by the tidal field of the Galaxy.
A key element of the scenario is that $\omega$ Cen was formed from
a GMC within $\sim 300$ Myr,  which is much shorter than those assumed
in previous studies.
We have derived chemical abundances of various elements
(e.g., He, C, N, O, Mg, Al, Ba, La)
from the simulated GC
and the spatial variations of these within the GC
(e.g., radial gradient of chemical abundances).
We have briefly discussed how the results can change,
if we adopt different yield tables of AGB stars
and SNe II from other groups (e.g., V13).
The principal results are as follows:

(1)  After the bursty formation of first generation (1G) of stars from the
original GMC,
second generation (2G) of stars can be later formed from gas that is ejected 
from 1G AGB stars. 
A significant fraction of 2G stars 
can have higher He abundances ($Y$) and higher
[Fe/H], because AGB ejecta from 1G can be polluted by SNe II yet can be retained
in the central region of the GC's dwarf for secondary star formation.
The He-rich (2G) stars are more strongly
concentrated in the central region of the GC in comparison with He-poor
(1G) stars, and this result does not depend
on model parameters.
The simulated $\omega$ Cen initially  has a flattened shape and
the He-rich population  ($Y>0.25$) 
shows a slightly more flattened shape than  the He-poor
population ($Y<0.25$).
The simulated $\omega$ Cen has 
a negative radial gradient of $Y$
with the slope of $-0.003$ pc$^{-1}$ within the central 10 pc,
and this result does not depend on model parameters.

(2) The  simulated $\omega$ Cen shows a small amplitude of 
rotation ($<5$ km s$^{-1}$) in its He-poor population, and the
rotational amplitude depends strongly on the initial ratio of
rotational energy to total kinetic energy in its natal GMC.
The observed rotational amplitude of $\omega$ Cen  can be better reproduced
by the models with $f_{\rm rot} \sim 0.03$ for the adopted virial ratios
of GMCs. He-rich population  show a  less amount of rotation, because 
it is formed from gas mixed with SNe II ejecta. The smaller amplitude of
rotation in He-rich (mostly 2G) stars in the present study is therefore
in a striking contrast with previous simulations in which 
2G stars are formed  from AGB ejecta without mixing with
SNe II ejecta.

(3) The GC stars have bimodal number distributions
in [C/Fe], [N/Fe], [O/Fe], and [Na/Fe]
owing to the formation of 2G stars 
from AGB ejecta. For example,
new stars formed from AGB ejecta mixed with SNe II ejecta
have rather low [O/Fe] ($<0$) and higher [Fe/H] 
whereas those formed from original gas polluted by 
SNe II ejecta (that is not well mixed
with AGB ejecta) have higher [O/Fe] and higher [Fe/H].
This bimodal [O/Fe] distribution in the simulated GC can be seen in
the observed [O/Fe] distribution for $\omega$ Cen (e.g., JP10).
The simulated $\omega$ Cen does not show clear bimodal distributions
in [Mg/Fe] and [Al/Fe].

(4) Negative radial gradients (i.e., higher abundances for inner regions)
of [Fe/H], [C/Fe] and [N/Fe] 
can be seen in the simulated $\omega$ Cen,
however, such gradients cannot be clearly seen in [O/Fe], [Na/Fe],
[Mg/Fe] and  [Al/Fe].
The simulated negative [Fe/H] gradient is not clearly observed
(JP10), which implies that the long-term dynamical evolution of the GC
should wipe out the original gradient to explain the observation.
There is a weak positive radial 
gradient for [O/Fe] in some models
of the present study. The lower [O/Fe] in the central region
of the GC is due to the accumulation of massive AGB ejecta in the central
region where new stars can be formed. Since these positive or negative
radial abundance gradients are not so  strong even in the early phase
of $\omega$ Cen formation, they could be wiped out over $\sim 10$ Gyr due to
the long-term two-body relaxation processes of $\omega$ Cen.

(5) The distribution of stars on the [N/Fe]-$Y$ plane 
(``chromosome map'') for simulated $\omega$ Cen
shows a wide range of [N/Fe]  from $\sim 0$
to $\sim 1.5$ for each of the three major
populations with different $Y$.  This is mainly because AGB ejecta with
high $Y$ and high [N/Fe] can mix with SNe II ejecta with high $Y$ and 
very low [N/Fe] ($\sim -0.8$) to different degrees in different local
star-forming regions.
The distributions of stars on the [O/Fe]$-$[Fe/H] and  [O/Na]$-$[Fe/H] planes for
simulated $\omega$ Cen have a bimodality, though they
are rather complicated.
However the details of the distributions  are not
so consistent with corresponding observations owing to the presence of
a large number of stars with large [O/Fe] ($>0.5$), as long as K06 is adopted
for [O/Fe] in SNe II.
These [O/Fe]$-$[Fe/H] and [O/Na]$-$[Fe/H] relations
can be changed, if AGB yields from other groups
(e.g., V13) are incorporated into the present simulations,
because [O/Fe] in stellar winds of  AGB stars is  different
between different groups. 
The lower [O/Fe] predicted for massive AGB stars in V13 implies
that the relations can be better reproduced, if their AGB yields are adopted
in the present study.

(6) The chemical abundances of $s$-process elements, [La/Fe] and [Ba/Fe],
rapidly increase around [Fe/H] $\sim -1.5$ 
in the early formation phase of $\omega$ Cen
owing to  star formation
from gas ejected from AGB stars. 
However, [La/Fe] and [Ba/Fe] soon decrease
with increasing [Fe/H], mainly because 2G SNe II can heavily pollute the 
intra-cluster medium (ICM) from which new stars are formed. 
The simulated $\omega$ Cen shows rather weak negative gradient of
[La/Fe] and [Ba/Fe], in particular, within the central few pc.
The [La/Fe] and [Ba/Fe] number distributions show  weak second peaks around
[La/Fe] and [Ba/Fe]$\approx$0.5, though the bimodal distributions are not so clear as
[O/Fe] and [Na/Fe].
The efficient trapping of AGB ejecta and the partial retention of NSM ejecta
cooperate to increase [La/Eu] from $-0.6$ to 0.6
for $-1.7 < {\rm [Fe/H]} < -1.5$. This difference in the retention capabilities
between AGB and NSM ejecta is the essential factor for the observed
[La/Fe]$-$[Fe/H] relation in $\omega$ Cen, because full retention of NSM ejecta
can dramatically decrease [La/Eu].

(7) NSMs both from 1G and 2G stars play a significant role in the evolution
of [Eu/Fe], because NSM ejecta can be partially trapped and retained in
ICM
of the forming GC owing to the rather high gas density 
of the ICM.  Ejection of $r$-process elements (e.g, Eu) can temporarily
increase [Eu/H] to a large extent within the early formation
phase of the GC.
2G SNe II can decrease [Eu/Fe]
of the ICM due to Fe ejection whereas  NSMs can increase [Eu/Fe] later. Therefore,
the time evolution of [Eu/Fe] depends strongly on (i) the formation
rates of SNe II and NSMs and (ii) the retention
capability of these ejecta by ICM  in the GC.
The fiducial model shows a decreasing [Eu/Fe] with increasing [Fe/H],
which is consistent with some observations (e.g., D11)
yet inconsistent with JP10 showing 
an almost constant [Eu/Fe]  over
a wide range of [Fe/H].
The present study can reproduce well the presence 
of some stars with unusually low [Eu/Fe] ($<0$)
observed in $\omega$ Cen.

(8) The moderately  wide range of [Fe/H] observed in $\omega$ 
Cen can be reproduced
only if the IMF of 2G stars is top-light (i.e., a smaller number of
massive stars that explode as  SNe II). If a canonical IMF is adopted for
2G stars, then the most metal-rich stars can be too metal-rich
([Fe/H]$>0$),
though the number is small.
Although the total mass of He-rich stars in the simulated $\omega$ Cen
($\sim 7 \times 10^5 {\rm M}_{\odot}$)
is consistent with observation,
the mass fraction of He-rich and metal-rich
2G stars  
($<0.1$)
is  smaller 
than the observed fraction ($\approx 0.2$).
This problem can be mitigated if the total mass of 1G 
stars becomes much smaller owing to (i) selective
stripping of 1G stars from the GC during its merging with the Galaxy
or (ii) the top-heavy IMF of 1G stars.

(9) The observed flattened shape of $\omega$ Cen can be due to the angular
momentum of its natal GMC formed in the central region of its host 
dwarf galaxy. The formation process of such a massive GMC in the dwarf
is beyond the scope of this paper. The global rotation of the GMC is 
responsible for the formation of 1G population with 
a significant amplitude of rotation,
and such rotation of 1G population and SNe II feedback effects
combine to control the rotational amplitude 
of 2G population in the GC. It is our future study to investigate 
the long-term dynamical evolution 
of these initially rotating populations of the GC orbiting the Galaxy.

(10) Although the present model can explain at least qualitatively
some of
the observed characteristic features in the chemical abundance patters
of $\omega$ Cen, the level of consistency
between observations and simulations  is far from satisfactory. For example, 
the simulated  distributions of stars on the
[N/Fe]$-$Y,  [O/Fe]$-$[Fe/H], and  [O/Na]$-$[Fe/H]
maps are not so consistent with observations.
Also, it is found to be very  difficult
for the present simulation
to identify  $\sim 15$ distinct subpopulations
observed in $\omega$ Cen, because
the formation of 2G stars in the simulation is not discrete.
These problems will need to be addressed in our future simulations with
more sophisticated modeling of mixing of SNe II, AGB, and NSM ejecta
and metallicity-dependent chemical yields and stellar evolution
(e.g., evolution of He-rich AGB stars).

(11) Thus, a number of unique physical processes in $\omega$ Cen formation
can be responsible for the observed properties that are distinguished from
those of other ordinary GCs with multiple stellar populations yet without
[Fe/H] spreads. $\omega$ Cen is the crossroad of astrophysical processes
in the sense that its unique properties reflect various physical processes
involved in its formation:
Table 6 presents a brief summary of these processes.
For example,  
the partial retention of NSM ejecta and efficient mixing
of SNe II, AGB, and NSM ejecta cooperate  to cause 
the observed unusually low [Eu/Fe] ($<0$) at [Fe/H]$>-1$ 
and rather high [La/Eu] at [Fe/H] $\sim -1.5$ in
$\omega$ Cen.
Its chemical enrichment due predominantly to SNe II (not SNe Ia) over
a short period ($\sim 300$ Myr) has something to do with the unique
formation processes of stars (e.g., top-light IMF and the longer minimum
delay time of SNe Ia).
Given that the present study has not successfully
explained a few of the observed properties of $\omega$ Cen so well,
some new physical processes will need to be included in our
future simulations of $\omega$ Cen formation.

\acknowledgments
We are  grateful to the referee  for  constructive and
useful comments that improved this paper.

\begin{deluxetable}{llllllllll}
\footnotesize  
\tablecaption{ Description of the  values of major parameters
for the representative models.
\label{tbl-1}}
\tablewidth{-2pt}
\tablehead{
\colhead{  ID  \tablenotemark{a}} &
\colhead{  $M_{\rm gmc}$  \tablenotemark{b}} &
\colhead{  $R_{\rm gmc}$ \tablenotemark{c}} &
\colhead{  $f_{\rm rot}$ \tablenotemark{d}} &
\colhead{  $r_{\rm gmc}$  \tablenotemark{e}} &
\colhead{  Yield  \tablenotemark{f}} &
\colhead{  Mixing  \tablenotemark{g}} &
\colhead{  $r_{\rm nsm}$  \tablenotemark{h}} &
\colhead{  $l_{\rm s,0}$  \tablenotemark{i}} &
\colhead{  $f_{\rm ret}$  \tablenotemark{j}}  }
\startdata
M1 & 2 & 140 & 0.01 & 0 & Y3 & D & 2.7 & 0.78 & --  \\
M2 & 2 & 140 & 0.01 & 0 & Y1 & D & 2.7 & 0.78 & --  \\
M3 & 2 & 140 & 0.01 & 0 & Y2 & D & 2.7 & 0.78 & --  \\
M4 & 2 & 140 & 0.01 & 0 & Y4 & D & 2.7 & 0.78 & --  \\
M5 & 2 & 140 & 0.01 & 0 & Y1 & D & 2.7 & 0.26 & --  \\
M6 & 2 & 140 & 0.01 & 0 & Y1 & D & 2.7 & 2.6 & --  \\
M7 & 2 & 140 & 0.01 & 0 & Y1 & D & 2.7 & 26.0 & --  \\
M8 & 2 & 140 & 0.01 & 0 & Y1 & D & 0.27 & 26.0 & --  \\
M9 & 2 & 140 & 0.01 & 0 & Y1 & U & 5.0 & -- & 0.01  \\
M10 & 2 & 140 & 0.01 & 0 & Y4 & U & 5.0 & -- & 0.01  \\
M11 & 2 & 140 & 0.01 & 0 & Y1 & U & 30.0 & -- & 1.0  \\
M12 & 2 & 140 & 0.01 & 0 & Y1 & U & 5.0 & -- & 1.0  \\
M13 & 2 & 140 & 0.03 & 0 & Y3 & D & 2.7 & 0.78 & --  \\
M14 & 2 & 140 & 0.05 & 0 & Y3 & D & 2.7 & 0.78 & --  \\
M15 & 2 & 280 & 0.01 & 0 & Y3 & D & 2.7 & 0.78 & --  \\
M16 & 2 & 140 & 0.01 & 40 & Y3 & D & 2.7 & 0.78 & --  \\
M17 & 2 & 140 & 0.01 & 100 & Y3 & D & 2.7 & 0.78 & --  \\
M18 & 2 & 140 & 0.01 & 200 & Y3 & D & 2.7 & 0.78 & --  \\
M19 & 2 & 140 & 0.01 & 500 & Y3 & D & 2.7 & 0.78 & --  \\
M20 & 2 & 140 & 0.01 & 1000 & Y3 & D & 2.7 & 0.78 & --  \\
M21 & 1 & 140 & 0.01 & 0 & Y3 & D & 2.7 & 0.78 & --  \\
M22 & 6 & 140 & 0.01 & 0 & Y3 & D & 2.7 & 0.78 & --  \\
\enddata
\tablenotetext{a}{
M1 corresponds to the fiducial model.
}
\tablenotetext{b}{
The initial total mass of a  GMC in units of $10^7 M_{\odot}$.
}
\tablenotetext{c}{
The initial radius of a GMC in units of pc.
}
\tablenotetext{d}{
The initial fraction of rotational energy among
total kinetic energy in a GMC.
}
\tablenotetext{e}{
The initial position of a GMC from the center of its host dwarf
galaxy in units of pc.
}
\tablenotetext{f}{
Yield models for AGB ejecta. The yield table is given in Table 2.
}
\tablenotetext{g}{
Mixing model for NSM ejecta. ``D'' and ``U'' represent the density-dependent
and uniform models, respectively.
}
\tablenotetext{h}{
The search radius (pc) for neighboring gas particles around a NSM event.
}
\tablenotetext{i}{
The stopping length (pc) for the reference gas density ($10^5$ atom cm$^{-3}$).
}
\tablenotetext{j}{
The retention fraction of NSM ejecta for the uniform mixing model.
}
\end{deluxetable}

\newpage
\begin{deluxetable}{lll}
\footnotesize  
\tablecaption{Observational diagnostics for 
the present new model of $\omega$ Cen with a short formation
timescale ($\sim 300$ Myr).
If observations  are well reproduced by the present model,
 ``\checkmark '' are given.  If they are not  reproduced by the present model,
and if there are promising ways to resolve this inconsistency,
``$\triangle$'' are given.
If observations are not reproduced by the present model, and if
there are no promising solutions for them, ``$\times$'' are given.
\label{tbl-1}}
\tablewidth{0pt} 
\tablehead{
}
\startdata
Item  &  Consistency & Comments  \\
Flattened shape & \checkmark &  \\
Global rotation  & \checkmark &  Higher $f_{\rm rot}$ required \\
Large [Fe/H] dispersion  & \checkmark &  \\
He-rich, metal-rich population  & \checkmark &  \\
Populations with different $Y$  & \checkmark &  \\
Radial gradient of $Y$  & \checkmark &  \\
Bimodal [O/Fe] distribution  & \checkmark &  \\
Fraction of O-poor stars  & \checkmark  &  \\
Central concentration of O-poor stars  & \checkmark  &  \\
Wide abundance ranges for C, N and Na  & \checkmark &  \\
Stellar distribution on [N/Fe]$-Y$   & $\triangle$  &   \\
Stellar distribution on [O/Na]$-$[Fe/H]  & $\times$  
& Some stars with [O/Na]$>0$    \\
Relation between [Ba/Fe] and [Fe/H]    & $\triangle$ &  
Peak at lower [Fe/H] ($\sim -1.5$) \\
Relation between [La/Fe] and [Fe/H]    & $\triangle$ & 
Peak at lower [Fe/H] ($\sim -1.5$)  \\
Relation between [Eu/Fe] and [Fe/H]   & \checkmark  &  \\
Higher [La/Eu] at [Fe/H]$\sim -0.5$   & \checkmark  & 
Larger La yield required \\
Low [Rb/Zr]  & $\times$  & Shorter lifetimes of He-rich stars ?  \\
Distinct 15 subpopulations  & $\times$  &   \\
\enddata
\end{deluxetable}

\newpage
\begin{deluxetable}{ll}
\footnotesize  
\tablecaption{A brief summary for the unique physical processes
in $\omega$ Cen formation that are responsible for its observed characteristics.
These processes that are investigated in this work and will be done in
future works are separately listed in this table.
\label{tbl-1}}
\tablewidth{0pt} 
\tablehead{
}
\startdata
Physical process  &  Relevant observed properties  \\
Present work  &    \\
(1) Retention and mixing of SNe II ejecta in ICM  & 
A wide range of [Fe/H] \\
(2) Star formation from AGB ejecta mixed with SNe II  & 
He-rich populations with higher [Fe/H] \\
(3) 2G formation preferentially in the central region & 
Negative radial gradient of $Y$ \\
(4) Top-light IMF in 2G formation  & 
Efficient 2G formation  \\
(5) No/little dilution of AGB ejecta by ISM & 
Stars with very low [O/Na] ($<-0.6$)  \\
(6) Full (partial) retention of AGB (NSM) ejecta  & 
Sharp rise of [La/Eu] from [Fe/H]$\sim -1.6$ \\
(7) No/little contribution of SNe Ia  & 
No/little evolution of $[\alpha$/Fe]  \\
(8) Partial retention of NSM ejecta  & 
Unusually low [Eu/Fe] ($<0$) \\
(9) Global rotation of the natal GMC  & 
Rotation of Y-poor stars \\
Future work  &    \\
(10) Star formation truncation by gas stripping  & 
Short formation timescale of $\omega$ Cen \\
(11) Time-evolving IMF in 2G formation  & 
Zone of avoidance in [Na/O]-[Fe/H] map \\
(12) Evolution of He-rich AGB stars  & 
Low [Rb/Zr] \\
(13) Discrete star formation events   & 
15 subpopulations \\
(14) Formation of massive GMC in a dwarf   & 
The origin of its unusually large mass  \\
(15) Accretion of its host  dwarf   & 
Retrograde orbit around the Galaxy  \\

\enddata
\end{deluxetable}

\newpage
\begin{figure}
\epsscale{1.0}
\plotone{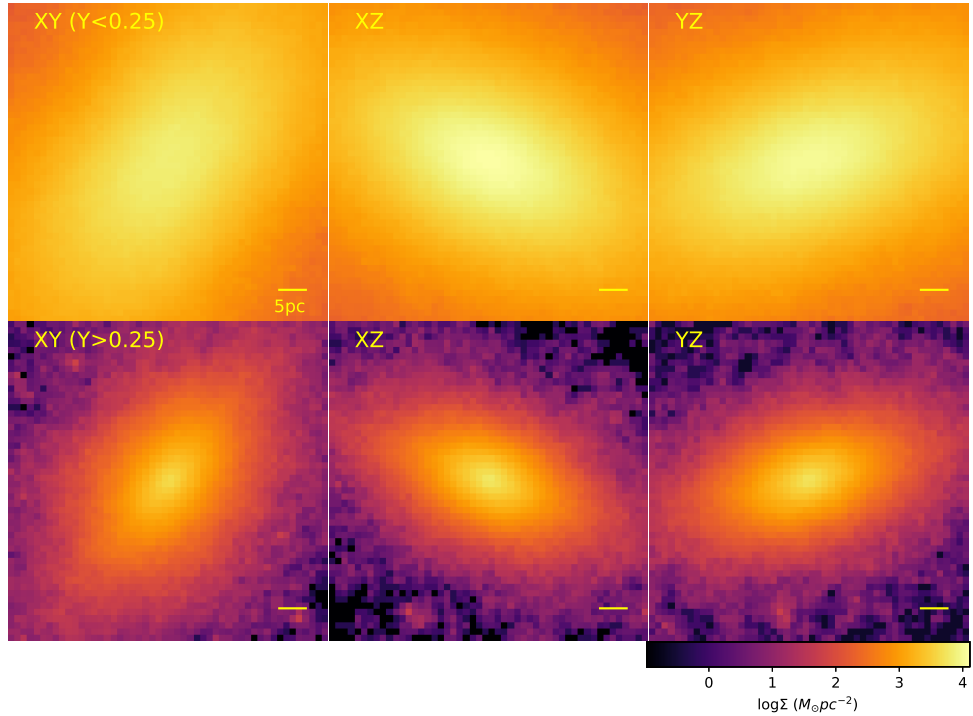}
\figcaption{
Mass distributions of ``He-poor'' ($Y \le 0.25$,
upper) and ``He-rich'' ($Y>0.25$, lower)
stars projected onto to the $x$-$y$ (left), $x$-$z$ (middle), and
$y$-$z$ (right) planes for the simulated $\omega$ Cen at the final time step
($T=298$ Myr) in the fiducial model M1. The surface mass density
is estimated at each mesh point and shown in these 2D maps with $50 \times 50$ meshes.
\label{fig-1}}
\end{figure}

\newpage
\begin{figure}
\epsscale{1.0}
\plotone{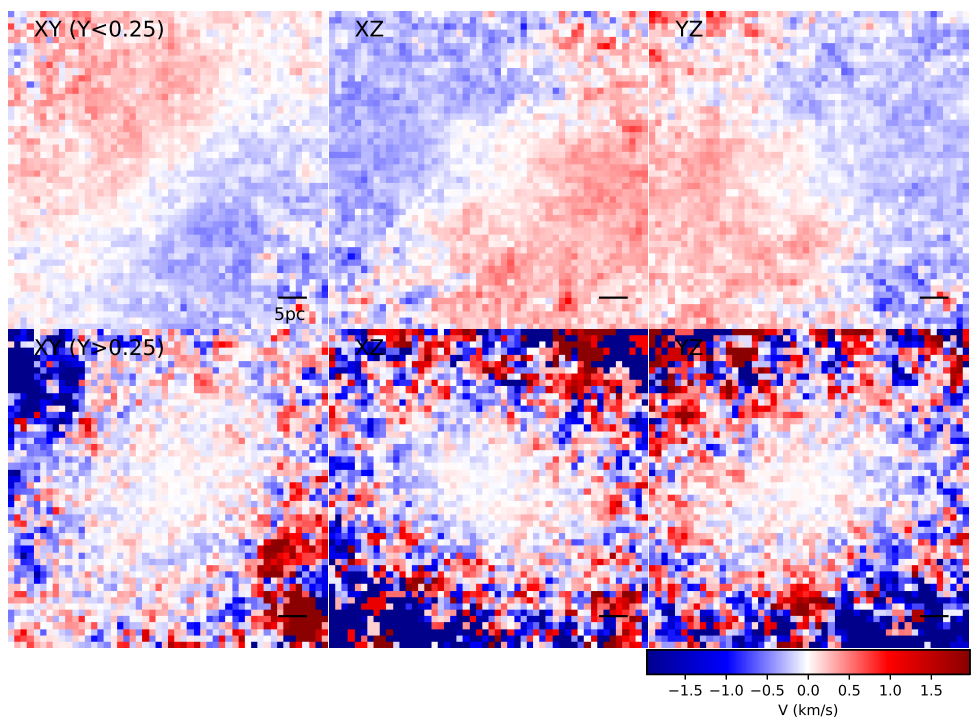}
\figcaption{
2D kinematics of He-poor ($Y \le 0.25$)
and He-rich ($Y>0.25$)  stars for the three projections
at the final time step
in the fiducial model M1.
The 2D map is based on the line-of-sight velocities ($V_{\rm los}$) of 
stars within 30pc.
For clarity, meshes with no He-rich particles are given the minimum value
of $V_{\rm los}$.
Also, meshed with $V_{\rm los} >2$ km s$^{-1}$ or $<-2$ km s$^{-1}$
are given reddest and bluest colors, respectively, so that the color contrast
between different regions can be more clearly seen.
\label{fig-2}}
\end{figure}

\newpage
\begin{figure}
\epsscale{1.0}
\plotone{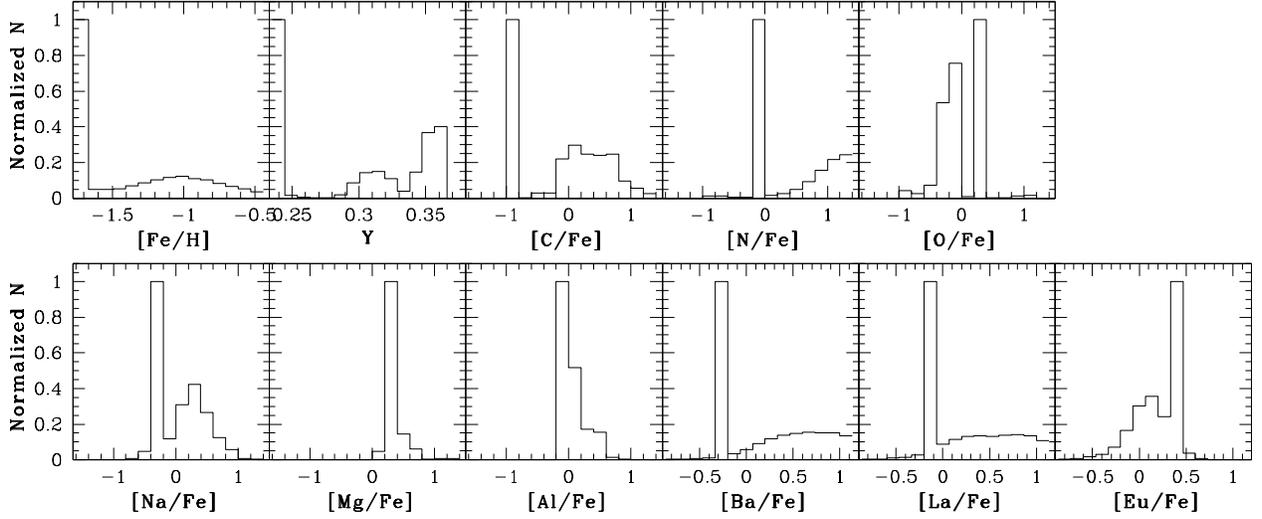}
\figcaption{
Abundance distribution functions (ADFs) of the 11 elements for
the simulated $\omega$ Cen  at the final time step in the fiducial model.
For clarity, the number of stars in each bin is normalized by the maximum
number of stars in the bins for each abundance. 
Only 30\% of 1G stars are used for this analysis so that the ADFs of 1G and 2G stars
can be clearly seen: if all 1G stars are used, then the distributions of 2G stars
cannot be clearly seen owing to the dominant 1G population. As discussed in the main text,
1G stars are more efficiently stripped by the Galaxy during tidal interaction of
$\omega$ Cen with the Galaxy owing to its more diffuse distribution. Therefore,
removing a significant fraction of 1G stars in this analysis is physically reasonable.
The location of the peak in the [C/Fe] distribution ($\sim -1$) reflects
the adopted low initial carbon abundances of gas particles in the host 
GMC in this model.
Therefore, this peak is not due to chemical enrichment processes during 
the GC formation (e.g., ejection of C-poor ejecta from AGB stars into
ICM).
\label{fig-3}}
\end{figure}

\newpage
\begin{figure}
\epsscale{1.0}
\plotone{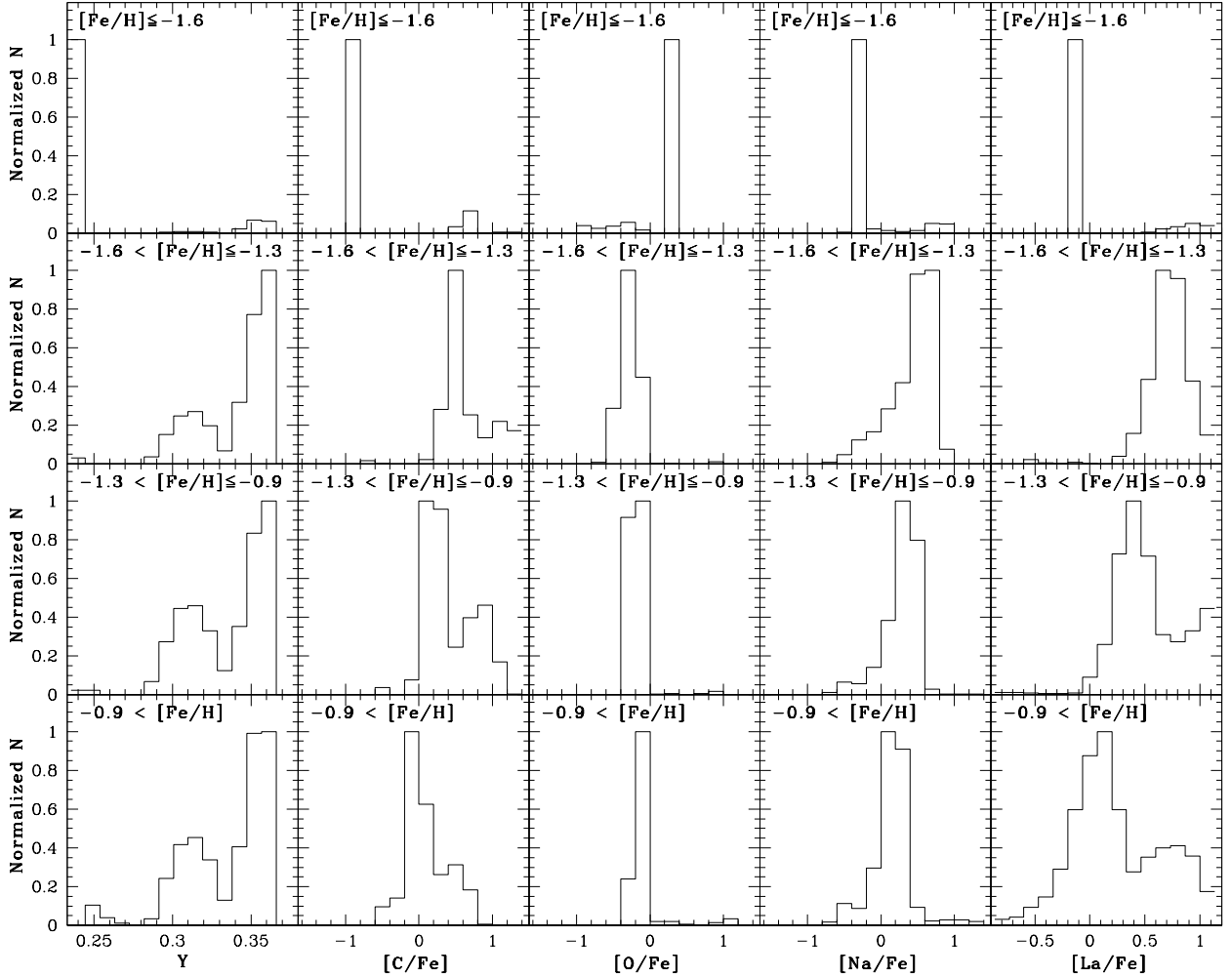}
\figcaption{
Abundance distribution functions of $Y$, [O/Fe], [Na/Fe], and [La/Fe]
for four different metallicity ranges 
in the simulated $\omega$ Cen  at the final time step in the fiducial model M1.
The metallicity ranges are chosen such that these results can be compared with
observations by JP10.
\label{fig-4}}
\end{figure}

\newpage
\begin{figure}
\epsscale{0.8}
\plotone{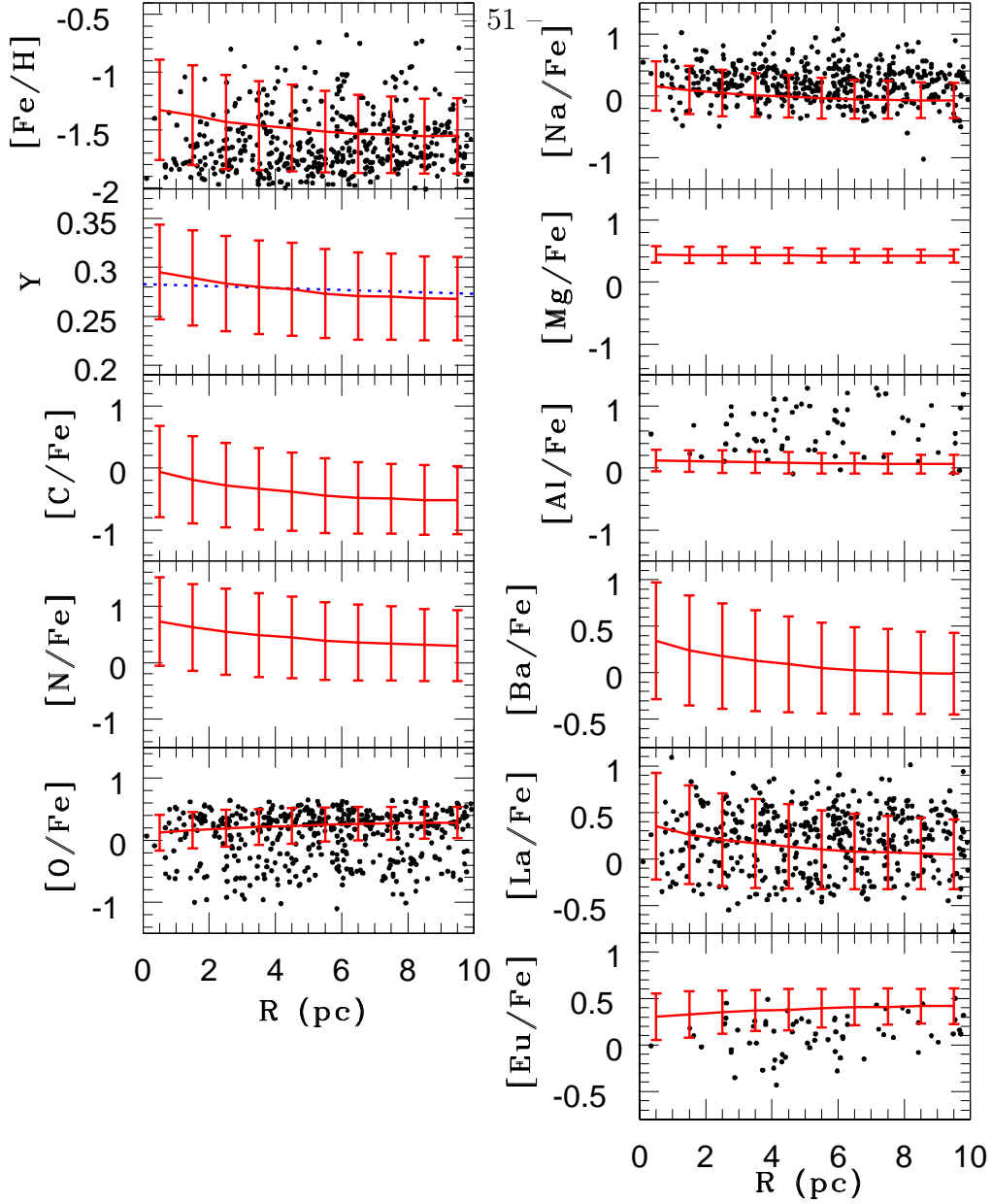}
\figcaption{
Radial gradients of the 11 elements 
in the simulated $\omega$ Cen  at the final time step in the fiducial model M1
(red lines).
The error bar shown in each radial bin for each
abundance  indicate the $1\sigma$ dispersion of the abundance.
Black filled circles represent the observational results from JP10
and the blue dotted line for the radial gradient of $Y$ is estimated
from the observed radial profile of the number ratio ($R_{\rm bMS}$)
of blue main-sequence
stars (``bMS'') to red main-sequence stars (``rMS'') by  Sollima et al. 
(2007). Here $Y$ is assumed to be 0.25 for rMS and 0.35 for bMS.
The $Y$ gradient is negative yet shallow (only 0.01 difference over 10 pc),
though $f_{\rm bms}$ changes from 0.48 at $R=0$ pc to 0.29 at $R=10$ pc.
\label{fig-5}}
\end{figure}


\newpage
\begin{figure}
\epsscale{0.8}
\plotone{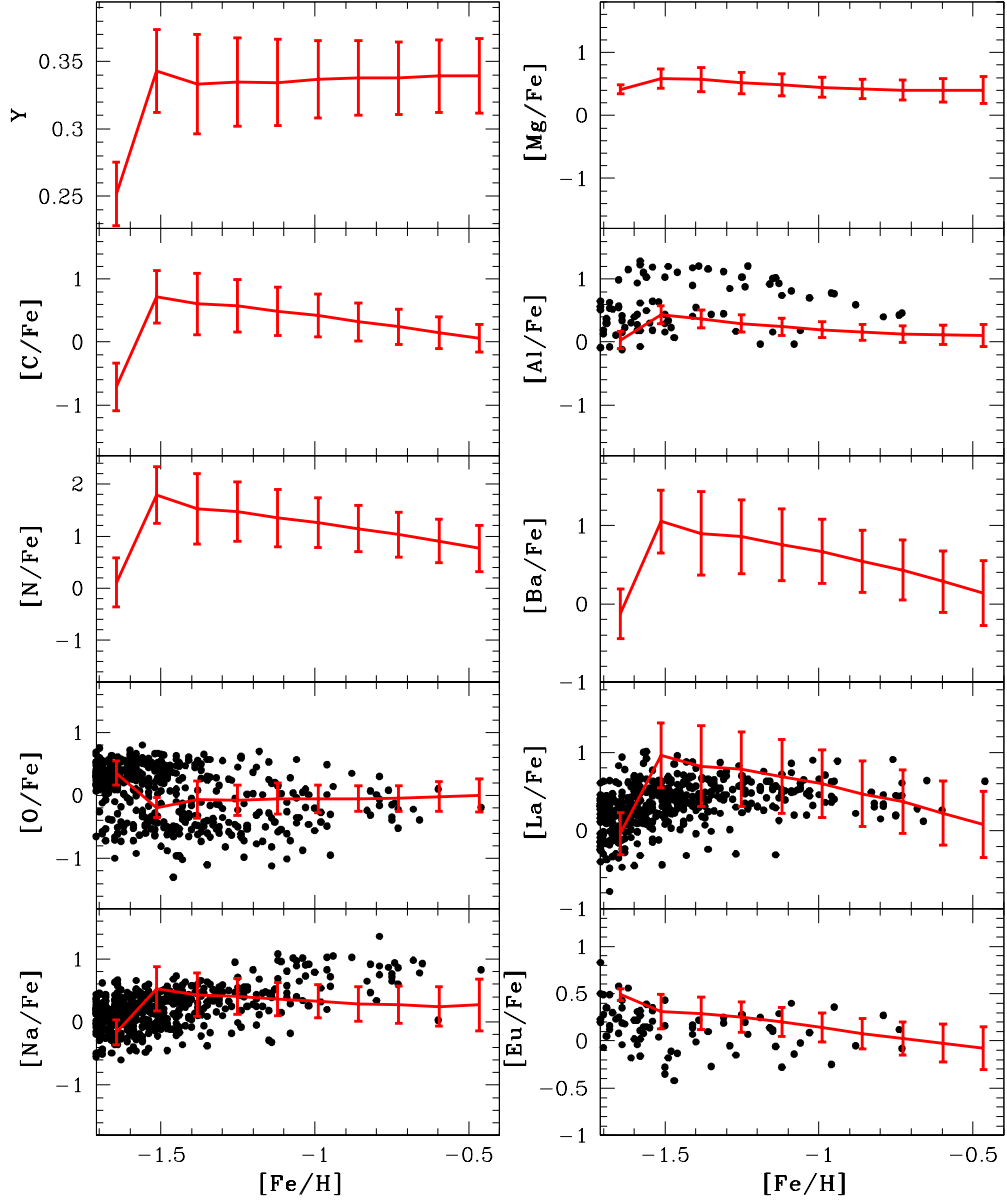}
\figcaption{
Relations between $Y$ and [Fe/H] and between ${\rm [X/Fe]}$ and [Fe/H],
where X= C, N, O, Na, Mg, Al, Ba, La, and Eu,
in the simulated $\omega$ Cen in the fiducial model M1
(red lines).
The error bar shown in each [Fe/H] bin for each
abundance  indicate the $1\sigma$ dispersion of the abundance.
Black filled circles represent the observational results from JP10.
\label{fig-6}}
\end{figure}

\newpage
\begin{figure}
\epsscale{0.7}
\plotone{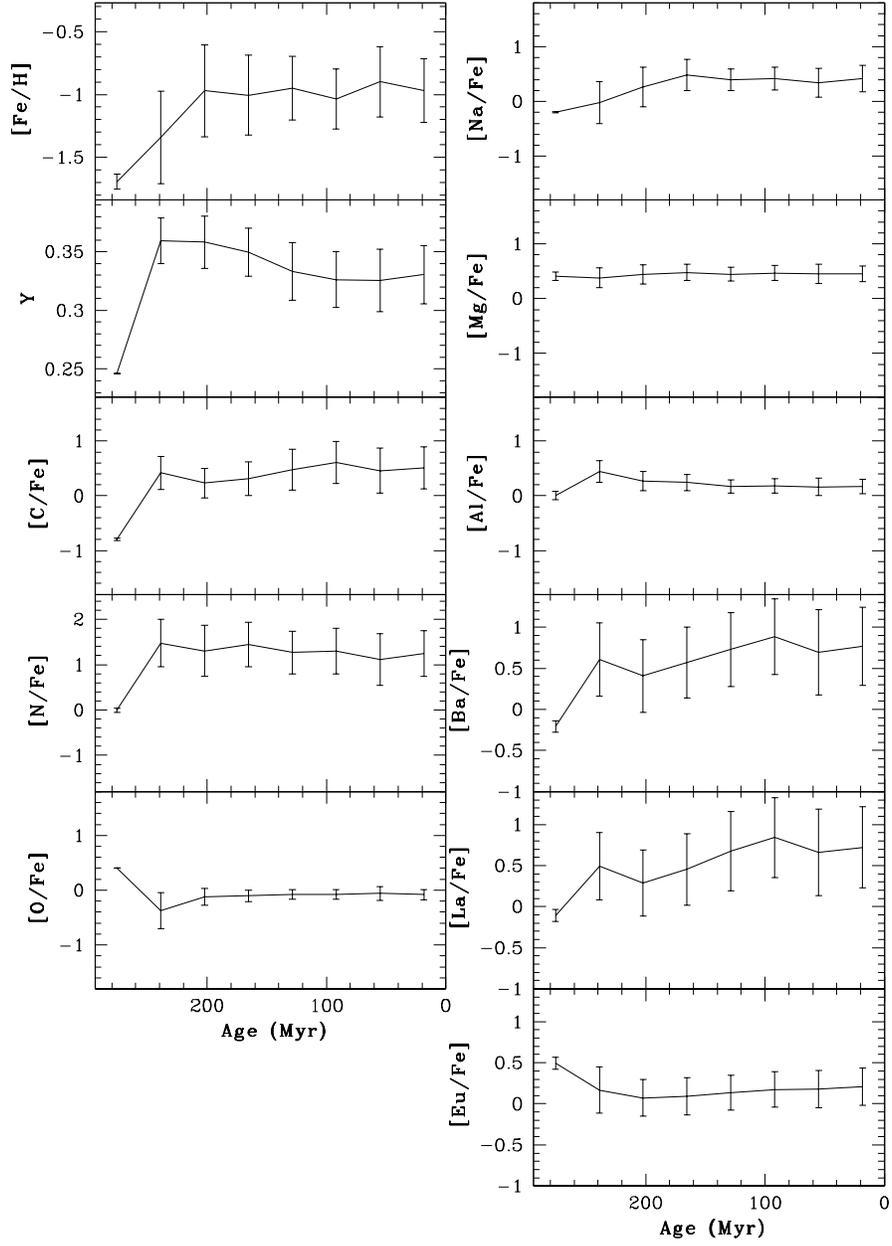}
\figcaption{
The age-abundance relation for the 11 elements
in the simulated $\omega$ Cen in the fiducial model M1.
The error bar shown in each age bin for each
abundance  indicate the $1\sigma$ dispersion of the abundance.
\label{fig-7}}
\end{figure}

\newpage
\begin{figure}
\epsscale{0.8}
\plotone{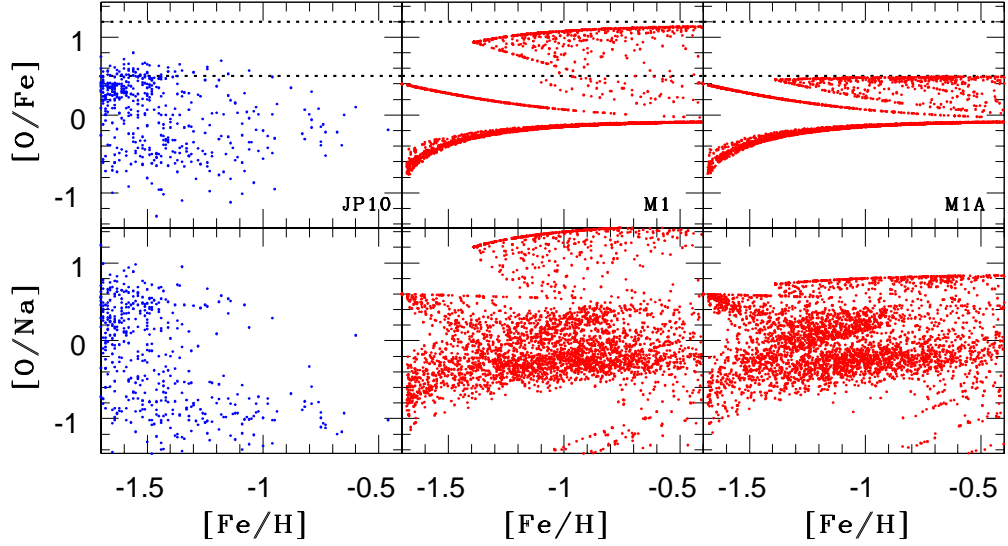}
\figcaption{
A comparison between the observed (blue, left) and simulated distributions (red, middle
and right)
of stars 
on the [O/Fe]$-$[Fe/H] (upper two) and [O/Na]$-$[Fe/H] planes (lower two).
The fiducial model M1 and the comparative model M1A
in which model parameters 
are exactly
the same as M1 except for
[O/Fe] in the ejecta of massive SNe II.
For convenience, only $\approx 4000$ stars from simulated
are selected and plotted.
The adopted [O/Fe] for massive SNe II 
are set to be 0.5 for $m \ge 15 M_{\odot}$ in M1A.
The  yields of [O/Fe]  from K06 and Tominaga et al. (2007)
are shown by upper and lower dotted lines, respectively, so that the results
can be more clearly understood. The significant fraction  of stars
with high [O/Fe] ($>1$) for [Fe/H]$>-1.0$) is due to the adoption of yields
from K06 in the present study.
\label{fig-8}}
\end{figure}

\newpage
\begin{figure}
\epsscale{0.8}
\plotone{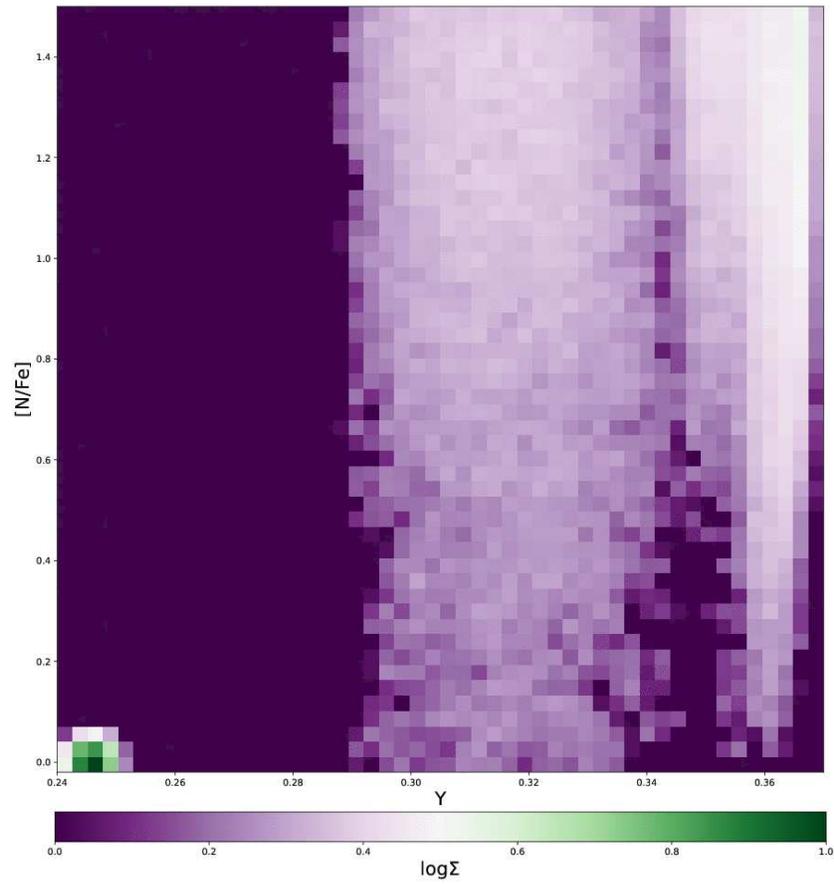}
\figcaption{
Distribution of stars 
on the $Y$-[N/Fe] plane in the fiducial model.
For clarity, the number density is normalized so that it can range from 0 to 1
in logarithmic scale. The highest density part at the lower left corner corresponds
to the dominant 1G population. Three populations with different $Y$ can be
clearly seen in this map.
\label{fig-9}}
\end{figure}

\newpage
\begin{figure}
\epsscale{0.8}
\plotone{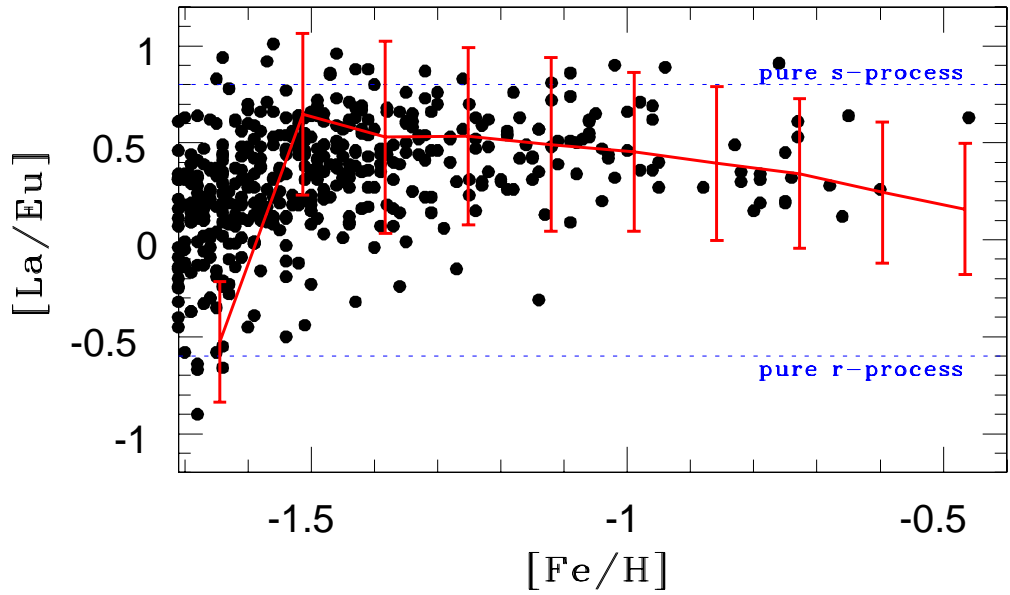}
\figcaption{
Evolution of [La/Eu] with [Fe/H] for the fiducial model,
which can be compared with corresponding results from JP10 (e.g., their Figure 18).
The abundance ratios for pure $r-$ and $s$-process enrichment 
(Simmerer et al. 2004) are shown by dotted lines.
Black filled circles represent the observational results from JP10.
\label{fig-10}}
\end{figure}

\newpage
\begin{figure}
\epsscale{1.0}
\plotone{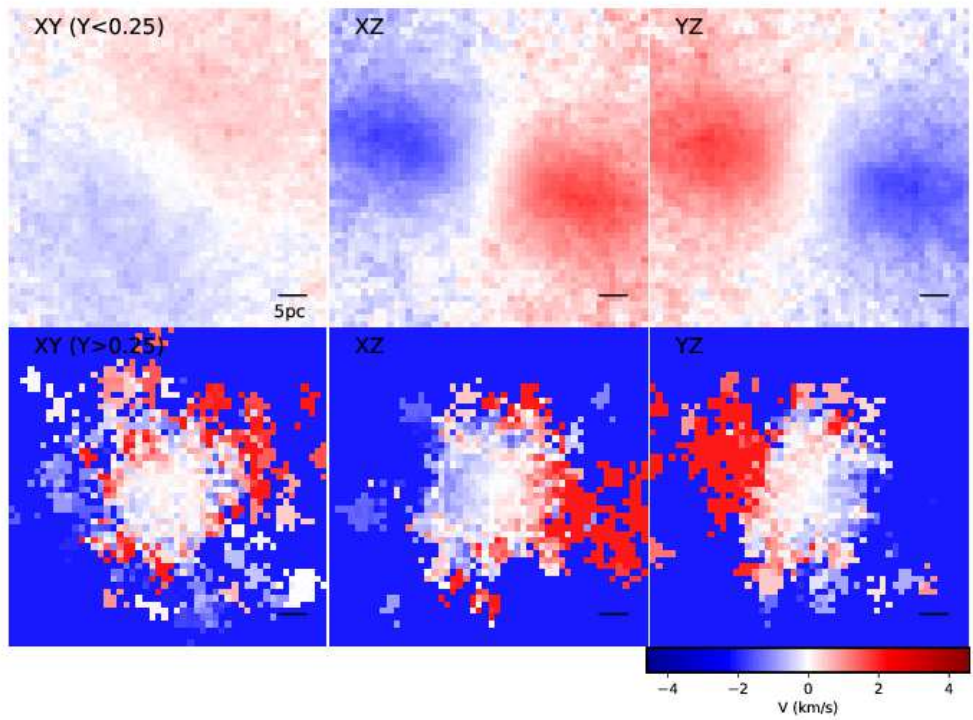}
\figcaption{
The same as Figure 2 but for the model M2 with $f_{\rm rot}=0.03$ 
(more initial angular momentum of the natal GMC).
\label{fig-11}}
\end{figure}

\newpage
\begin{figure}
\epsscale{0.8}
\plotone{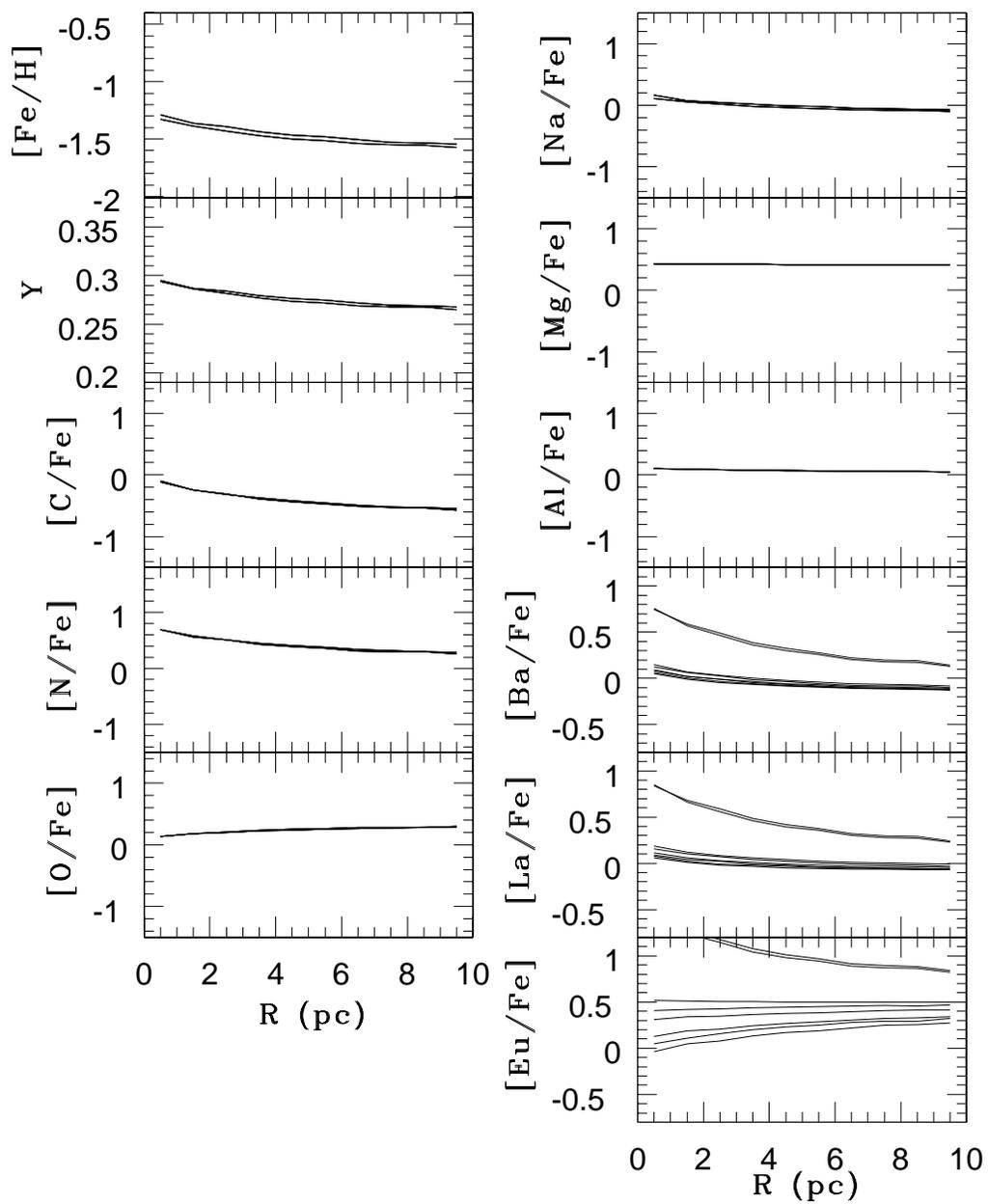}
\figcaption{
Radial gradients of the 11 elements for the representative 8 models
with different mixing models (M2, M5, M6, M7, M8, M9,  M11, and  M12).
\label{fig-12}}
\end{figure}

\begin{figure}
\epsscale{1.0}
\plotone{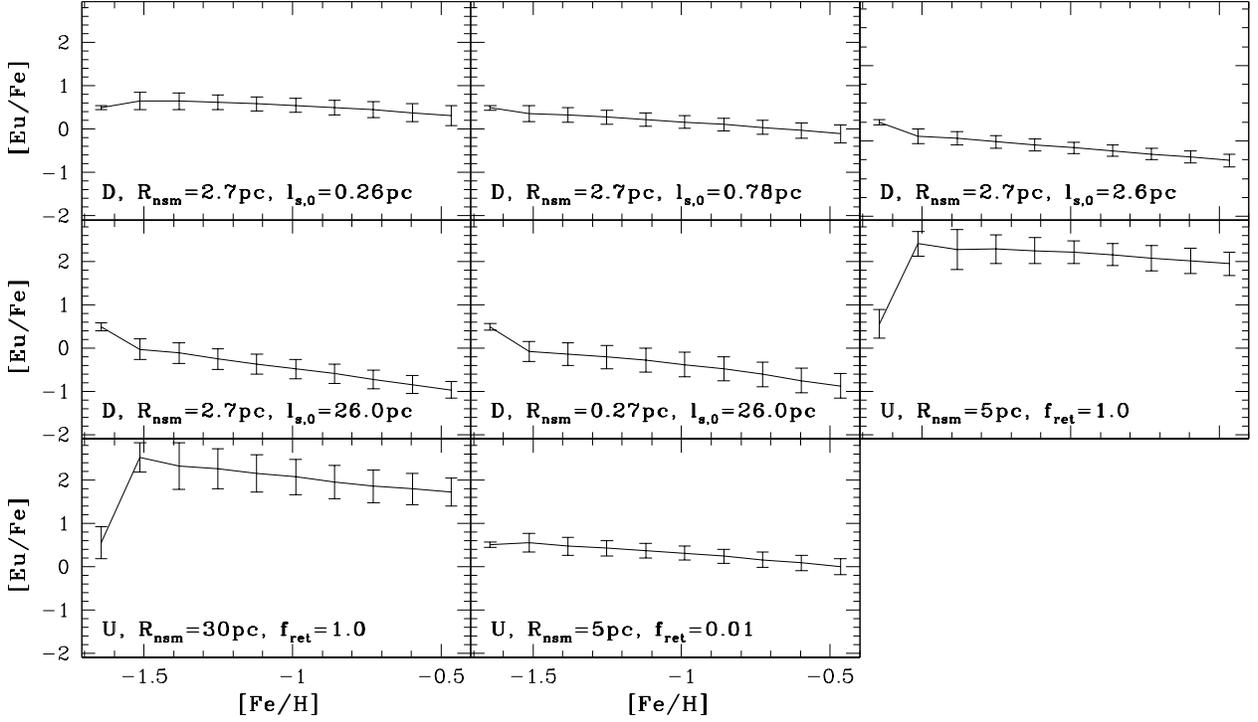}
\figcaption{
Evolution of [Eu/Fe] with [Fe/H] for 8 different mixing 
models (``D'' or ``U'')
of NSM ejecta  with different $r_{\rm nsm}$ and $l_{\rm s, 0}$
(M2, M5, M6, M7, M8, M9,  M11, and  M12).
\label{fig-13}}
\end{figure}

\begin{figure}
\plotone{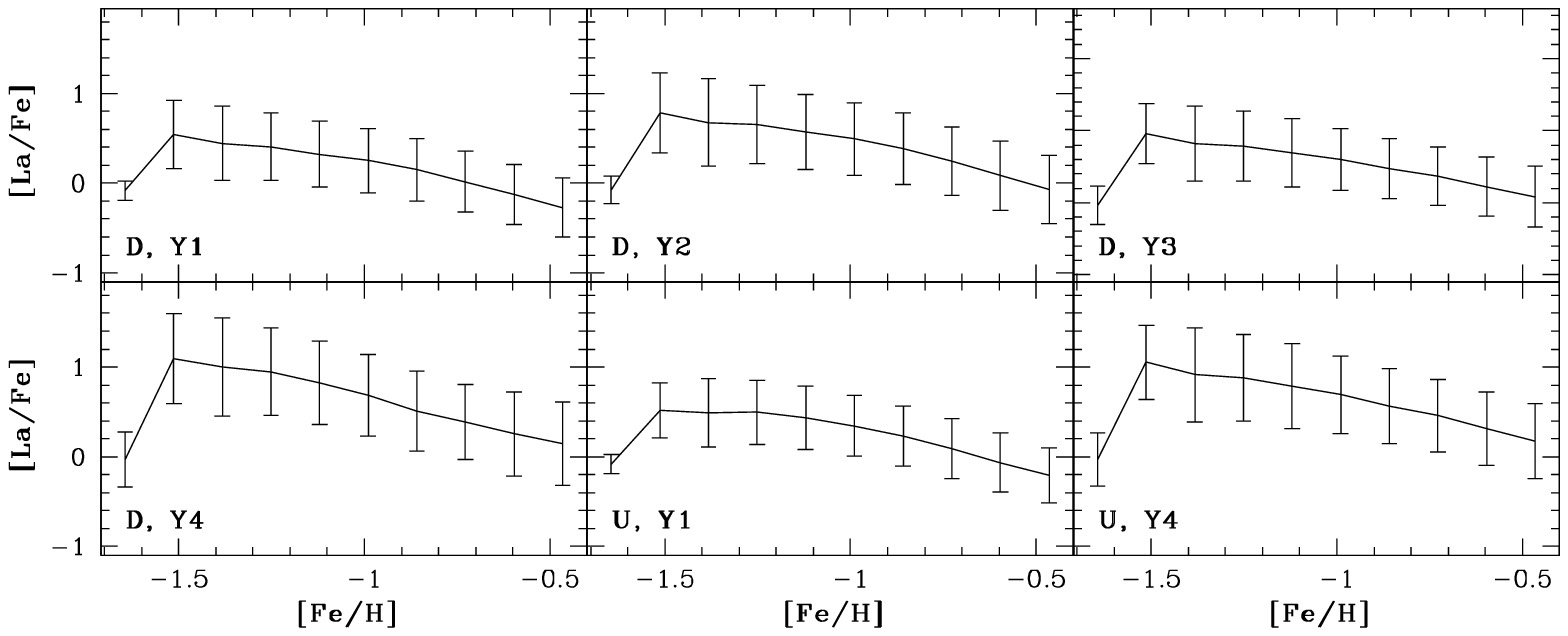}
\figcaption{
Evolution of [La/Fe] with [Fe/H] for 6 different mixing
models with different AGB yields
(M1, M2, M3, M4, M9, and  M10).
\label{fig-14}}
\end{figure}


\begin{thebibliography}{}


\bibitem[]{}
Anderson, J., 1997, PhD Thesis, UC, Berkeley

\bibitem[]{}
Bedin, L. R., 2004, ApJL, 605, 125

\bibitem[]{}
Bastian, N.,  Lardo, C., 2018, ARA\&A, 56, 83

\bibitem[]{}
Bekki, K., 2006, MNRAS, 367, L24

\bibitem[]{}
Bekki, K., 2011, MNRAS, MNRAS, 412, 2241 

\bibitem[]{}
Bekki, K., 2013, 432, 2298 (B13a)

\bibitem[]{}
Bekki, K., 2013, 436, 2254 (B13b)


\bibitem[]{}
Bekki, K., 2017a, MNRAS, 467, 1857 (B17a)

\bibitem[]{}
Bekki, K., 2017b, MNRAS, 469, 2933  (B17b)

\bibitem[]{}
Bekki, K., 2019, MNRAS, 486, 2570 (B19a)

\bibitem[]{}
Bekki, K., 2019, A\&A, 622, 53 (B19b)

\bibitem[]{}
Bekki, K., Freeman, K. C., 2003, MNRAS, 346, L11 (BF03)


\bibitem[]{}
Bekki, K.,  Chiba, M., 2007, ApJ, 665, 1164

\bibitem[]{}
Bekki, K., Campbell, S. W., Lattanzio, J. C., Norris, J. E.,
2007, MNRAS, 377, 335 (B07)

\bibitem[]{}
Bekki, K., Yong, D., 2012, MNRAS, 419, 2063

\bibitem[]{}
Bekki, K., Tsujimoto, T., 2016, ApJ, 831, 70 

\bibitem[]{}
Bekki, K., Tsujimoto, T., 2017, ApJ, 844, 34 (BT17)

\bibitem[]{}
Bekki, K., Jerabkova, T., Kroupa, P., 2017, MNRAS, 471, 2242 

\bibitem[]{}
Bergin, Edwin A.; Tafalla, M., 2008, ARA\&A, 2007, 45, 339

\bibitem[]{}
Bellini, A., et al., 2017, ApJ, 842, 7

\bibitem[]{}
Bellini, A., et al., 2018, ApJ, 853, 86

\bibitem[]{}
Bianchini, P., et al. 2013, ApJ, 772, 67


\bibitem[]{}
Bono, G., et al. 2019, ApJ, 870, 115

\bibitem[]{}
Busso, M., et al., 2001, ApJ, 557, 802


\bibitem[]{}
Carraro, G.,  Lia, C., 2000, A\&A, 357, 977

\bibitem[]{}
Cunha, K., et al., 2010,  ApJ, 717, 333

\bibitem[]{}
D'Antona, F., \&  Caloi, V. 2004, ApJ, 611, 871

\bibitem[]{}
D'Antona, F., et al., 2011, ApJ, 736, 5

\bibitem[]{}
D'Ercole, A., Vesperini, E., D'Antona, F., McMillan, S. L. W.,
\& Recchi, S. 2008, MNRAS, 391, 825 


\bibitem[]{}
Dinescu, D. I., et al. 1999, AJ, 117, 277

\bibitem[]{}
D'Orazi, V., et al. 2011, A\&A, 534, 29 (D11)

\bibitem[]{}
Dominik et al. 2012, ApJ, 759, 52

\bibitem[]{}
Ferraro, F., et al., 2004, ApJL, 603, 81


\bibitem[]{}
Fishlock, C. K., et al., 2014, ApJ, 797, 44


\bibitem[]{}
Freeman, K.,  \& Rodgers, A. W.,  1975, ApJ, 201, 71

\bibitem[]{}
Freeman, K., 1993, in  Smith  G.  H.,  Brodie  J.  P.,  eds,  ASP  
Conf.  Ser.Vol. 48, The Globular 
Clusters-Galaxy Connection. Astron. Soc. Pac.,San Francisco, p. 608

\bibitem[]{}
Gnedin, O. Y., et al.,  1999, 522, 935

\bibitem[]{}
Gratton, R. G., et al. 2011, A\&A, 534, 72




\bibitem[]{}
Hilker, M., Richtler, T., 2000, A\&A, 362, 895

\bibitem[]{}
Hughes, J., Wallerstein, G., 2000, AJ, 119, 1225

\bibitem[]{}
Ibata, R. A., et al., 2019, NatAs, 258

\bibitem[]{}
Ideta, et al., 2004, ApJL, 616, 107


\bibitem[]{}
Ikuta, C.,  Arimoto, N., 2000, A\&A, 358, 535

\bibitem[]{}
Jindal, et al., 2019, preprint (arXiv190311070)

\bibitem[]{}
Johnson, C. I., Pilachowski, C. A., 2010, ApJ, 722, 1373 (JP10)

\bibitem[]{}
Karakas, A. I., et al.,  2014, ApJ, 784, 32

\bibitem[]{}
Karakas, A. I., et al.,  2018, MNRAS, 477, 421

\bibitem[]{}
Kobayashi, C., et al. 2006, ApJ, 653, 1145 (K06)


\bibitem[]{}
Lee, Y.-W., et al., 1999, Nature, 402, 55

\bibitem[]{}
Lee,  Y.-W., et al. 2002,
Extragalactic Star Clusters, IAU Symposium 207, 
Eited by D. Geisler, E.K. Grebel, and D. Minniti. San Francisco.,
 Astronomical Society of the Pacific,  p.110


\bibitem[]{}
Magurno, D., et al. 2019, preprint, arXiv:1906.08550

\bibitem[]{}
MaKenzie, M., Bekki, K., 2018, MNRAS, 479, 3126

\bibitem[]{}
Marcolini, A., et al., 2007, MNRAS, 382, 443

\bibitem[]{}
Marino, A. F., Milone, A. P., Piotto, G., Villanova, S., Bedin, L. R.,
Bellini, A., Renzini, A.,
2009, A\&A, 505, 1099

\bibitem[]{}
Marino, A. F., et al., 2011, ApJ, 731, 64

\bibitem[]{}
Marino, a., et al. 2019, in preprint (arXiv190405180)

\bibitem[]{}
Marks, M., et al. 2012, MNRAS, 422, 2246

\bibitem[]{}
Mastrobuono-Battisti, A.,  Perets, H. B., 
2016, ApJ, 823, 61

\bibitem[]{}
Meylan,  G., 1987, A\&A, 184, 144

\bibitem[]{}
Meylan, G., Mayor, M., Duquennoy, A., Dubath, P., 1995, A\&A, 303, 761




\bibitem[]{}
Mizutani, A., et al, 2003, ApJL, 589, 89

\bibitem[]{}
Myeong, G. C., et al., 2018, MNRAS, 478, 5449

\bibitem[]{}
Norris, J. E., 2004, ApJL, 610, 25

\bibitem[]{}
Norris, J. E.,  \& Da Costa, G. S.,  1995, ApJ, 447, 680

\bibitem[]{}
Origlia, L., et al. 2003, ApJ., 591, 916

\bibitem[]{}
Pancino, E., et al., 2002, ApJL, 568, 101

\bibitem[]{}
Pancino, E., et al., 2011, A\&A, 527, 18

\bibitem[]{}
Phillips, J. P., 1999, A\&AS, 134, 241

\bibitem[]{}
Piotto, G., et al., 2005, 621, 777

\bibitem[]{}
Prantzos, N., Charbonnel, C., 2006, A\&A, 458, 135

\bibitem[]{}
Renzini, A., et al., 2015, MNRAS, 454, 4197

\bibitem[]{}
Rey, S-C., et al. 2000,  AJ, 119,  1824

\bibitem[]{}
Romano, D., et al., 2007, MNRAS, 376, 405

\bibitem[]{}
Romano, D., et al., 2011, MNRAS, 418, 696

\bibitem[]{}
Rosolowsky, E., Engargiola, G.,  Plambeck, R., Blitz, L.,
2003, ApJ, 599, 258




\bibitem[]{}
Siegel, D. M., Barnes, J., Metzger, B. D.,  2019, Nature, 569, 241

\bibitem[]{}
Simmerer, J., et al. 2004, ApJ, 617, 1091


\bibitem[]{}
Smith, V., et al. 2000, AJ, 119, 1239 (S00)

\bibitem[]{}
Sollima, A., et al., 2005, MNRAS, 357, 265

\bibitem[]{}
Sollima, A., et al., 2007, ApJ, 654, 915

\bibitem[]{}
Stanford, L. M., et al. 2006, ApJ, 647, 1075

\bibitem[]{}
Sutherland R. S., Dopita M. A.
1993, ApJS 1993, 88, 2531


\bibitem[]{}
Tominaga, N., Umeda, H., Nomoto, K., 2007, ApJ, 660, 516

\bibitem[]{}
Totani, T., et al. 2008, PASJ, 60, 1327

\bibitem[]{}
Tsujimoto, T., \&  Shigeyama, T., 2003, ApJ, 590, 803

\bibitem[]{}
Tsujimoto, T., \&  Bekki, K., 2012, ApJ, 747, 125

\bibitem[]{}
Tsujimoto, T., \&  Shigeyama, T., 
2014, ApJ, 795, L18 (TS14)

\bibitem[]{}
Tsujimoto, T., Yokoyama, T., \&  Bekki, K. 2017, ApJ, 835, L3

\bibitem[]{}
Vanture, A. D., et al., 1994, PASP, 106, 835

\bibitem[]{}
Ventura, P., et al. 2013, MNRAS, 431, 3642 (V13)

\bibitem[]{}
Vesperini, E., McMillan, S. L. W.,  D'Antona, F., \& D'Ercole, A.
2010, ApJ, 718, L112

\bibitem[]{}
Villanova, S., et al. 2007, ApJ, 663, 296

\bibitem[]{}
Wylie de-Boer, E., et al., 2010, AJ, 139, 636



\end{thebibliography}
\end{document}